\documentclass[lettersize,journal]{IEEEtran}

\usepackage[T1]{fontenc}
\usepackage{amsmath,amssymb}
\usepackage{cite}

\usepackage{booktabs}
\usepackage{graphicx}
\usepackage{array}
\usepackage{multirow}
\usepackage{colortbl}
\usepackage[most]{tcolorbox}
\usepackage{listings}
\usepackage{fontawesome5}
\usepackage[normalem]{ulem}
\usepackage{enumitem}
\usepackage{longtable}
\usepackage{balance}

\usepackage{microtype}
\usepackage{xurl}
\usepackage[hidelinks]{hyperref}

\definecolor{rq3task}{HTML}{E7F4F2}
\definecolor{rq3output}{HTML}{EFE9F6}
\definecolor{rq3constraint}{HTML}{F8E8E7}
\definecolor{rq3process}{HTML}{F7EEDB}
\definecolor{rq3context}{HTML}{E8F0F8}
\definecolor{rq3resource}{HTML}{E6F1F4}
\definecolor{rq3example}{HTML}{F3EEDD}
\definecolor{rq3communication}{HTML}{F5E8EF}
\definecolor{rq3evaluation}{HTML}{E6F3EA}
\definecolor{rq3safety}{HTML}{F7E6E4}
\definecolor{rq3category}{HTML}{D7E8F2}
\definecolor{rq4best}{HTML}{E5F3E7}
\definecolor{idebg}{HTML}{FFFFFF}
\definecolor{idefg}{HTML}{24292F}
\definecolor{idekey}{HTML}{116329}
\definecolor{idevalue}{HTML}{0969DA}
\definecolor{idecomment}{HTML}{6E7781}
\definecolor{idemarker}{HTML}{CF222E}

\lstdefinestyle{rq3markdown}{
  basicstyle=\fontsize{8}{9.4}\selectfont\ttfamily,
  backgroundcolor=\color{gray!5},
  aboveskip=3pt,
  belowskip=3pt,
  breaklines=true,
  columns=fullflexible,
  keepspaces=true,
  showstringspaces=false
}

\lstdefinestyle{backgroundmarkdown}{
  basicstyle=\scriptsize\ttfamily,
  breaklines=true,
  columns=fullflexible,
  keepspaces=true,
  showstringspaces=false
}

\newtcblisting{workflowmdexample}{
  enhanced,
  listing only,
  colback=gray!4,
  colframe=black!55,
  boxrule=0.45pt,
  boxsep=0.5mm,
  arc=1mm,
  left=1.2mm,
  right=1.2mm,
  top=0.2mm,
  bottom=0.2mm,
  listing options={style=backgroundmarkdown}
}

\newtcolorbox{realworkflowmdexample}{
  enhanced,
  colback=idebg,
  colframe=black!30,
  coltext=idefg,
  boxrule=0.45pt,
  boxsep=0.8mm,
  arc=1mm,
  left=1.2mm,
  right=1.2mm,
  top=0.8mm,
  bottom=0.8mm
}

\newcommand{\yamlkey}[1]{\textcolor{idekey}{\texttt{#1}}}
\newcommand{\yamlvalue}[1]{\textcolor{idevalue}{\texttt{#1}}}
\newcommand{\yamlcomment}[1]{\textcolor{idecomment}{\texttt{#1}}}
\newcommand{\yamlmark}[1]{\textcolor{idemarker}{\texttt{#1}}}
\newcommand{\mdheadingtwo}[1]{\par\vspace{0.35em}{\large\bfseries\textcolor{black}{#1}}\par\vspace{0.1em}}

\newcommand{\rqmdinline}[1]{\textcolor{idevalue}{\texttt{#1}}}
\newcommand{\rqmdheading}[2]{\par\smallskip{\sffamily\bfseries
  \ifnum#1<2\fontsize{8.7}{10.2}\selectfont\else
  \ifnum#1<3\fontsize{8.2}{9.7}\selectfont\else
  \fontsize{7.7}{9.2}\selectfont\fi\fi #2\par}\nobreak\smallskip}

\newcommand{\RQone}{\textit{ RQ$_{1}$. What are the characteristics of GitHub Agentic Workflows?}}

\newcommand{\RQtwo}{\textit{ RQ$_{2}$. How are GitHub Agentic Workflows maintained over time?}}

\newcommand{\RQthree}{\textit{ RQ$_{3}$.What agentic instructions do developers set in GitHub Agentic Workflows?}}

\newcommand{\RQfour}{\textit{ RQ$_{4}$. Can LLMs assign instruction labels in GitHub Agentic Workflows?}}

\newcounter{rqsnippet}

\newtcolorbox{rq3mdexample}[5]{
  enhanced,
  colback=white,
  colframe=black!25,
  coltext=idefg,
  fontupper=\sffamily\fontsize{7.2}{8.8}\selectfont,
  before upper={\refstepcounter{rqsnippet}\label{#5}\raggedright},
  boxrule=0.4pt,
  boxsep=0.5mm,
  arc=1mm,
  left=2mm,
  right=2mm,
  top=1mm,
  bottom=1mm,
  before skip=0.55\baselineskip,
  after={\par\nobreak\vspace{2pt}\noindent
    \parbox{\linewidth}{\centering\fontsize{7.2}{8.5}\selectfont\sffamily
    Snippet~\ref*{#5}. #1.\\
    \href{#4}{\nolinkurl{#2/#3}}}\par
    \vspace{0.6\baselineskip}}
}

\newcommand{\rqpromptlabel}[1]{\par\addvspace{4pt}\noindent
  {\bfseries #1}\par\nobreak}
\newtcolorbox{rq4prompttemplate}{
  enhanced,
  title={LLM labeling task prompt},
  fonttitle=\rmfamily\fontsize{9}{11}\selectfont\bfseries,
  colbacktitle=white,
  coltitle=black,
  colback=white,
  colframe=black,
  coltext=black,
  fontupper=\rmfamily\fontsize{8.2}{10.3}\selectfont,
  before upper={\raggedright\setlength{\parindent}{0pt}},
  boxrule=0.4pt,
  titlerule=0.4pt,
  boxsep=0mm,
  sharp corners,
  left=2.5mm,
  right=2.5mm,
  top=1mm,
  bottom=2mm,
  toptitle=1.8mm,
  bottomtitle=1.8mm
}

\newcommand{\rqsubcat}[1]{\textbf{#1}}
\newcommand{\rqexamplesource}[1]{\textsuperscript{\href{#1}{\normalfont\scriptsize[source]}}}
\newcommand{\rqpart}[1]{\texorpdfstring{\textbf{\textit{\uline{#1}}}}{#1}}
\newcommand{\rqfinding}[2]{\noindent\textbf{Finding~\##1: #2}\par}
\newcommand{\threatlabel}[1]{\noindent\textbf{\textit{\uline{#1:}}}}
\newcommand{\rqthreelabel}[1]{}

\begin{document}

\title{Specifying and Maintaining Agentic Workflows: An Empirical Study of GitHub Agentic Workflows}

\author{Jasem~Khelifi, Issam~Oukhay, Ali~Ouni, Mohammed~Sayagh,
and~Mohamed~Aymen~Saied%
\thanks{J. Khelifi, I. Oukhay, A. Ouni, and M. Sayagh are with
\'{E}cole de technologie sup\'{e}rieure (\'{E}TS), University of Quebec,
Montreal, Quebec, Canada. E-mail: \nolinkurl{jasem.khelifi.1@ens.etsmtl.ca},
\nolinkurl{issam.oukhay.1@ens.etsmtl.ca}, \nolinkurl{ali.ouni@etsmtl.ca},
\nolinkurl{mohammed.sayagh@etsmtl.ca}.}%
\thanks{M. A. Saied is with Universit\'{e} Laval, Quebec City,
Quebec, Canada. E-mail: \nolinkurl{mohamed-aymen.saied@ift.ulaval.ca}.}%
\thanks{This work was supported by LLM token credits from the
Google Research Credits program.}}



\maketitle

\begin{abstract}
Agentic workflows shift software development from prompting AI agents for individual tasks to defining recurring work that agents execute automatically. GitHub Agentic Workflows (gh-aw) enables this approach through Markdown files that combine natural-language instructions with configuration and compile into executable GitHub Actions workflows. Unlike conventional workflows that primarily prescribe scripted operations, these files delegate tasks requiring interpretation to AI agents. They also couple agent instructions with execution triggers, making those instructions operational specifications for repeated repository activities. However, how developers structure and maintain these specifications, and which execution requirements and safeguards they express, remains insufficiently understood. 
In this paper, we examine the structure, evolution, and instruction content of gh-aw Markdown files to inform how practitioners define and maintain agent-run work.
We analyze 1,248 files from 276 repositories,
20,841 commit-file events, and 288 resolved instruction-label sets from a sample of 294 files.
Our results show that workflow instructions extend beyond short prompts, with a median of 556.5 words and code blocks in 62.1\% of files. Among files with at
least 120 days of observed activity, 78.2\% still receive updates in month 4, while size-normalized churn decreases after the first month. 
Tasks, outputs, constraints, and process
instructions each appear in over 93\% of labeled workflows, yet only 9.4\% explicitly
address prompt-injection defense. LLM classification achieves an F1 score of
0.818 and Cohen's \(\kappa\) of 0.715 against the resolved human labels.
These findings suggest that developers should account for the evolution
of workflows copied or referenced across repositories and consider adding
prompt-injection defenses, resource budgets, and evidence-credibility checks
where applicable

\end{abstract}


\section{Introduction}
\IEEEPARstart{A}{gentic} AI extends software assistance from generating answers to planning
and carrying out tasks through tools, including writing code, running tests,
and revising generated output \cite{yang2024sweagent,carlini2026compiler}.
Coding agents are already being adopted across GitHub projects
\cite{robbes2026agentic}.
Agentic workflows extend this model to recurring automation: developers
define work that agents execute in response to schedules or repository
events, without requiring a person to initiate and guide each run
\cite{github-gh-aw-overview}.
This shift toward recurring agent-driven work is echoed in Sundar Pichai's description of how Google's engineers are using agentic workflows:

\begin{quote}
 \textit{``We're now shifting to truly agentic workflows. Our engineers are orchestrating fully autonomous digital task forces, firing off agents and
accomplishing incredible things.''}
\end{quote}

\begin{flushright}
\begin{minipage}{0.95\columnwidth}
\small\raggedleft
Sundar Pichai, CEO, Google and Alphabet\\
April 22, 2026 \cite{pichai2026cloudnext}
\end{minipage}
\end{flushright}

GitHub Agentic Workflows (gh-aw) supports this model by allowing developers
to specify repository tasks in Markdown files that combine natural-language
instructions with execution configuration
\cite{github2026agentic-preview,github-gh-aw-overview}.
Its compiler translates these files into GitHub Actions YAML workflows
that invoke agents on schedules or repository events
\cite{github-gh-aw-compilation}.
Developers can use these workflows to investigate failing tests, review
changes, or update documentation \cite{github-gh-aw-overview}.
Where conventional workflows primarily prescribe scripted operations,
gh-aw allows developers to delegate tasks requiring interpretation and
reasoning while configuring when and how agents execute.

Agentic workflows make natural-language instructions persistent specifications for recurring repository operations. In gh-aw, developers combine instructions describing the agent's work with configuration specifying triggers, permissions, and tool access
\cite{github-gh-aw-overview,github-gh-aw-compilation}.
Workflow authors must specify what the agent should accomplish, which information it should consider, and what outputs and constraints should guide its execution \cite{github-gh-aw-overview}. Execution configuration and task instructions thus express complementary aspects of the delegated behavior. Understanding and reviewing these workflows requires examining both their configuration and their natural-language content.

Recurring execution makes the quality and maintenance of these instructions consequential. The same workflow can invoke an agent repeatedly on schedules or repository events
\cite{github-gh-aw-overview,github-gh-aw-compilation}.
An unclear requirement or outdated assumption may therefore affect successive runs, potentially producing unsuitable changes or additional review work. Understanding what developers specify and how they revise these instructions
can inform how such workflows are written and maintained.


Prior research has examined repository workflow automation from
several perspectives, including the maintenance of GitHub Actions
YAML workflows \cite{valenzuela2024hidden,rostami2026githubactions},
tool support for extracting GitHub Actions build metrics
\cite{khelifi2025ghaminer}, and the performance, failures, and
recovery of CI workflows in microservice-based systems
\cite{khelifi2026continuous}.
In parallel, studies of agent context and skill files have
characterized the structure, evolution, and content of
natural-language guidance
\cite{chatlatanagulchai2026agentreadmes,hong2026anatomy}.
More recently, GHAW-H has provided a dataset linking gh-aw
Markdown histories to compiled YAML snapshots
\cite{valenzuela2026ghawhistories}.
Although these efforts provide foundations for investigating
agentic automation, they leave open how developers maintain
Markdown sources that combine agent instructions with execution
configuration for recurring automation. In particular, further
empirical evidence is needed to characterize how maintenance
activity evolves after adoption, how edits are distributed between
configuration and instructions, and which operational requirements
and safeguards developers explicitly encode.

In this paper, we aim to address this gap through an empirical study of the structure, evolution, and instruction content of gh-aw Markdown workflows. Our objective is to understand how developers specify recurring agent tasks and maintain these specifications after adoption, providing an empirical foundation for workflow review and instruction-analysis tools. 
We analyze 1,248 gh-aw Markdown files from 276 GitHub repositories and manually examine a sample of 294 files. Our analysis characterizes workflow composition, distinguishes changes to execution configuration
from changes to agent instructions, and develops a taxonomy of instruction types. We also evaluate whether LLMs can identify these types to support instruction analysis at scale. The following research questions guide our study:

\begin{itemize}
  \item \textbf{\RQone}
  \item \textbf{\RQtwo}
  \item \textbf{\RQthree}
  \item \textbf{\RQfour}
\end{itemize}

Our results show that gh-aw workflows contain substantial instructions, with
a median of 556.5 words per file. Moreover, among files with at least 120 days
of observed activity, 78.2\% receive updates in month 4. Size-normalized churn
decreases after the first month, but maintenance's share of repository commits
shows no statistically significant month-to-month change.
Additionally, we identify ten instruction
categories. Tasks (97.6\%) and outputs (95.8\%) dominate, specifying what agents
should do and what results they should produce. In contrast, safety (25.0\%)
and communication (63.9\%) are the least common, covering protective
instructions and how agents should communicate. Finally, LLM classification
achieves an F1 score of 0.818 and Cohen's \(\kappa\) of 0.715 against the resolved human
labels, indicating substantial agreement \cite{landis1977measurement}.

\subsection{Novelty statement}

This study provides new empirical findings on how developers specify, maintain, and reuse gh-aw Markdown files. The main contributions of this paper can be summarized as follows:

\begin{itemize}
  \item We develop a taxonomy of instructions in gh-aw Markdown files, with ten categories and 42 subcategories. Based on manual analysis, it describes how developers define tasks, outputs, execution steps, constraints, and safeguards, with definitions and examples to support workflow review.
  \item We examine how developers maintain and reuse gh-aw Markdown files across projects. We distinguish changes to frontmatter configuration from changes to instruction bodies and track maintenance over four months, accounting for workflow size and repository activity.
  \item We characterize the size, structure, and composition of gh-aw Markdown files, including how developers combine instructions, code blocks, and template expressions.
  \item We provide a replication package to support independent
replications and further research on agentic workflows.

\end{itemize}

\subsection{Open Science} We provide a replication package containing our
dataset, scripts, labeling material, and experiments online
\cite{markdown-workflows-replication}, for future replications and extensions.

\section{Background}

GitHub Agentic Workflows (gh-aw) is a GitHub CLI extension and workflow authoring system for defining AI-powered repository automation in Markdown and running it through GitHub Actions \cite{github-gh-aw-repository}. In a gh-aw
workflow, the source artifact is not a conventional hand-written Actions YAML file. Instead, developers write a Markdown file, typically under \texttt{.github/workflows/}, that combines structured configuration with natural-language instructions for an AI agent. The gh-aw compiler then turns the Markdown source into a generated Actions workflow that can be triggered by standard GitHub events, schedules, or manual dispatches.

This source format gives gh-aw files a two-part structure. The YAML frontmatter
at the top specifies operational parameters such as triggers, permissions, model
or engine choices, and tool access. The Markdown body below the frontmatter
describes the task the agent should perform, the context it should inspect, the
steps it should follow, and the expected output. Figure~\ref{fig:background-md-example}
shows an example from our corpus. The example illustrates how executable
configuration and prose instructions coexist in one \texttt{.md} file: the
frontmatter decides when and with what authority the workflow runs, while the
body states the agent-facing triage policy.

\begin{figure}[t]
  \centering
\begin{realworkflowmdexample}
{\scriptsize\ttfamily
\yamlmark{-{}-{}-}\par
\yamlcomment{\# Auto-Triage Workflow - adds triage/accepted when Copilot is assigned to an issue}\par
\yamlkey{on}:\par
\hspace*{1em}\yamlkey{issues}:\par
\hspace*{2em}\yamlkey{types}: \yamlvalue{[assigned]}\par
\yamlkey{permissions}:\par
\hspace*{1em}\yamlkey{issues}: \yamlvalue{write}\par
\yamlkey{safe-outputs}:\par
\hspace*{1em}\yamlkey{report-failure-as-issue}: \yamlvalue{false}\par
\hspace*{1em}\yamlkey{noop}: \yamlvalue{false}\par
\hspace*{1em}\yamlkey{add-labels}:\par
\hspace*{2em}\yamlkey{max}: \yamlvalue{3}\par
\yamlmark{-{}-{}-}\par
}

\mdheadingtwo{Auto-Triage on Copilot Assignment}

{\small You monitor issues for Copilot assignment. When Copilot is assigned to
an issue, you add the \texttt{triage/accepted} label.\par}

\mdheadingtwo{When to Run}

{\small Only process when:}
\begin{itemize}[leftmargin=1.2em,itemsep=0pt,topsep=1pt]
\small
\item The assignee is \texttt{Copilot} (the GitHub Copilot coding agent)
\item The issue has the \texttt{ai-fix-requested} label
\end{itemize}

\mdheadingtwo{Your Task}

{\small When Copilot is assigned to an issue with \texttt{ai-fix-requested}:}
\begin{enumerate}[leftmargin=1.2em,itemsep=0pt,topsep=1pt]
\small
\item \textbf{Add the \texttt{triage/accepted} label} - This indicates the issue is accepted for work
\item \textbf{Add the \texttt{ai-processing} label} - This shows the issue is being actively worked on
\end{enumerate}

\mdheadingtwo{Do Nothing If}

\begin{itemize}[leftmargin=1.2em,itemsep=0pt,topsep=1pt]
\small
\item The assignee is NOT Copilot
\item The issue doesn't have \texttt{ai-fix-requested} label
\item The issue already has \texttt{triage/accepted} label
\end{itemize}

\mdheadingtwo{Important}

\begin{itemize}[leftmargin=1.2em,itemsep=0pt,topsep=1pt]
\small
\item Copilot's login is exactly \texttt{Copilot} (case-sensitive)
\item Do not post any comments - just add the labels silently
\item This workflow is triggered by the ``Assign to Copilot'' button in the GitHub UI
\end{itemize}
\end{realworkflowmdexample}
  \caption{Auto-triage workflow in \texttt{kubestellar/console}
  (\texttt{auto-triage.md}).\protect\footnotemark}
  \label{fig:background-md-example}
\end{figure}
\footnotetext{Commit-pinned source:
\href{https://github.com/kubestellar/console/blob/2432c7b78891238f33065321541e640690c7f669/.github/workflows/auto-triage.md}{workflow snapshot}.}

\subsection{Motivating Example}
\label{sec:motivating-example}

\begin{figure}[t]
  \centering
  \includegraphics[width=.94\columnwidth]{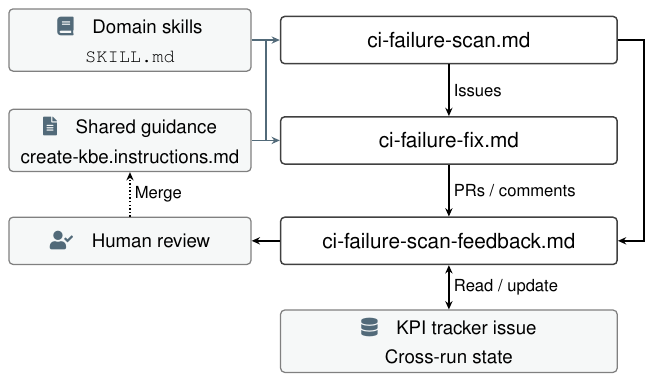}
  \caption{CI maintenance loop in \texttt{dotnet/runtime}.
  Selected information flows and reviewed updates to shared guidance.}
  \label{fig:runtime-motivating-example}
\end{figure}

To illustrate some of the key characteristics of agentic workflows,
we consider the CI maintenance workflows in
\texttt{dotnet/runtime},\footnote{\url{https://github.com/dotnet/runtime}}
the repository hosting the .NET runtime and core libraries.
Figure~\ref{fig:runtime-motivating-example} presents selected
information flows among three workflows and the human review
process for proposed updates to shared guidance.
Together, these workflows illustrate recurring execution,
coordination through repository artifacts, and maintenance
of agent instructions.

Every 12 hours, \texttt{ci-failure-scan.md} directs an agent
to investigate CI failures and file issues. A second workflow,
\texttt{ci-failure-fix.md}, reads those issues, consults
domain-specific skills, and attempts repairs. It opens draft
pull requests or asks the relevant owners for help when a fix
cannot be established
\cite{runtime2026scanner,runtime2026fixer}.
A third workflow, \texttt{ci-failure-scan-feedback.md}, runs daily
to assess both agents' outputs and maintainer feedback.
It proposes edits to their Markdown instructions and shared
issue-authoring guidance through a draft pull request
\cite{runtime2026feedback}.

The workflows operate on independent schedules and exchange
information through issues, pull requests, and run records.
The arrows in Figure~\ref{fig:runtime-motivating-example}
represent information flows rather than direct execution triggers.
Existing issues and pull requests help avoid repeated work,
while a KPI tracker issue preserves the feedback workflow's
tracking window across runs
\cite{runtime2026scanner,runtime2026fixer,runtime2026feedback}.
Proposed instruction changes affect subsequent runs after
review and integration.

A concrete incident illustrates why these instructions require
maintenance. A rejected attempt to modify a platform-specific
cryptography test led the feedback workflow to propose a rule
directing such cases to domain owners. After an automated push
failed, the proposal was preserved as an issue and subsequently
implemented in a reviewed, merged pull request
\cite{runtime2026feedbackissue,runtime2026feedbackfix}.
The resulting change revised the guidance available to future
agent runs, refining the boundary between tasks delegated to
agents and cases requiring domain expertise.

This example illustrates three concerns that motivate our study.
First, recurring agent work depends on both execution configuration
and natural-language instructions, including references to shared
guidance. Second, those instructions can evolve as experience
reveals limits in the tasks delegated to agents. Third, specifying
agent work involves defining the conditions under which agents
should act, refrain from acting, or seek human assistance.
Markdown therefore serves as a maintained specification of
delegated behavior and can itself become the target of
agent-proposed revisions subject to human review.

Although this example demonstrates these concerns in one project,
it does not establish how common they are across repositories.
Our study examines the structure of gh-aw Markdown workflows,
the evolution of their configuration and instructions, and the
requirements and constraints developers explicitly express.

\section{Study Design}
\label{sec:study-design}

\begin{figure*}[t]
  \centering
  \includegraphics[width=.8\textwidth,height=.29\textheight,keepaspectratio]{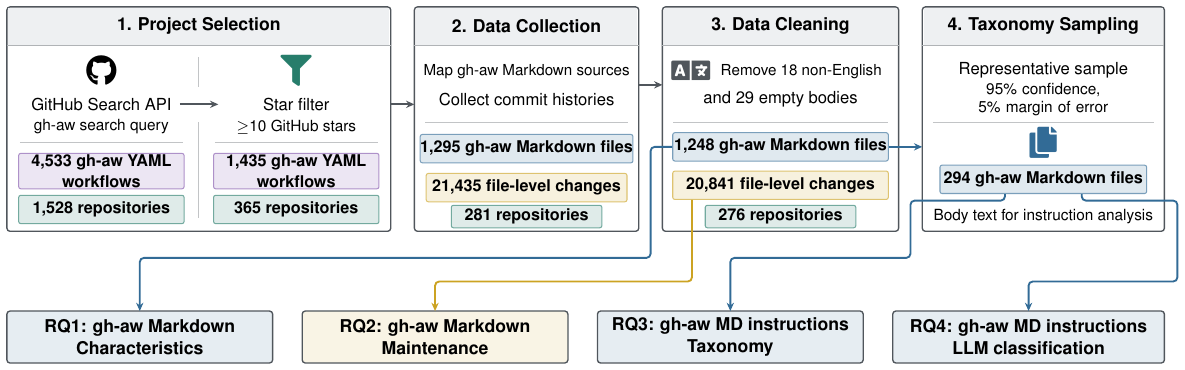}
  \caption{Overview of our study design.}
  \label{fig:study-design}
\end{figure*}

Figure~\ref{fig:study-design} outlines our study. We select projects
(Section~\ref{sec:study-selection}), collect Markdown sources and their commit
histories (Section~\ref{sec:study-collection}), clean the data
(Section~\ref{sec:study-cleaning}), and develop an instruction taxonomy
(Section~\ref{sec:study-taxonomy}).

\subsection{Project Selection}
\label{sec:study-selection}

We select projects by first identifying gh-aw generated GitHub Actions
workflow files through GitHub REST API code search \cite{github-rest-search} using the following search query:

\begin{tcolorbox}[colback=white,colframe=black!55,boxrule=0.4pt,
  sharp corners,boxsep=0mm,left=2mm,right=2mm,top=2mm,bottom=2mm,
  fontupper=\small\ttfamily,before upper={\raggedright}]
"This file was automatically generated\\
by gh-aw"\\
path:.github/workflows extension:yml
\end{tcolorbox}

GitHub code search allows at most 1,000 results per query across all pages
\cite{github-rest-search}. To retrieve more matches, we divide the query into
non-overlapping file-size buckets, such as \texttt{size:48001..60000} and
\texttt{size:60001..72000} (bytes). Each bucket has the same 1,000-result limit.
We repeatedly split buckets reporting more than this limit into smaller ranges.
All final buckets reported fewer than 1,000 matches. We then retrieve every 
page and removed duplicate repository/path pairs, targeting all indexed
gh-aw-generated YAML files matching the query rather than only the first 1,000.

After deduplicating all returned search results by repository and file path, we
identify 4,533 generated gh-aw workflow YAML files across 1,528 repositories.
We retain repositories with at least 10 GitHub stars as a proxy for
relevance and quality and to reduce the inclusion of toy projects, following
prior studies \cite{watanabe2026agenticprs,he2026cursor,salzano2026bridging}.
This left 365 repositories and 1,435 generated gh-aw workflow YAML files.

\subsection{Data Collection}
\label{sec:study-collection}

We map each YAML workflow to its Markdown source using its source comments
(e.g., \texttt{\# Source: owner/repo/path.md@sha}). When no source comment was
available, we infer the Markdown path in the same repository by replacing the YAML filename suffix with \texttt{.md} (e.g., \texttt{test.lock.yml} becomes
\texttt{test.md}). We deduplicate sources by repository and Markdown path,
since workflows in different repositories can reference the same Markdown file
hosted in another repository.
Starting from the 1,435 gh-aw generated YAML workflows, this mapping, deduplication,
and snapshot collection yield 1,295 distinct gh-aw Markdown source files
hosted by 281 repositories. Of the 365 projects with generated YAML workflows,
108 (29.6\%) reference Markdown from another repository for at least one
workflow, using a shared source rather than a project-local file to generate
its YAML.

For these Markdown files, we collect their commit histories,
yielding 7,653 distinct commits involving 21,435 changes to Markdown files.
For each change, we record the lines added and deleted,
file status, commit identifier, message, author, committer, and dates. We save
Markdown snapshots at the modifying commits, or pre-deletion versions for
deleted files.

\subsection{Data Cleaning}
\label{sec:study-cleaning}

We filter the 1,295 Markdown files by removing (\textit{i}) 29 files with empty body text, and (\textit{ii}) 18 files confidently classified as non-English. To identify
non-English files, we use \texttt{langdetect} \cite{danilak2021langdetect}, a Python library that detects language by comparing short character sequences
with stored language profiles. We apply it to the body text of the 1,295 latest snapshots, excluding YAML frontmatter and code blocks. Short or uncertain-language bodies are retained unless empty.
This leaves 1,248 files from 276 repositories. We retain the histories of these files, covering
7,446 distinct commits that touch at least one retained Markdown file and
20,841 file-level changes. Table~\ref{tab:data-collection-summary} summarizes
the resulting dataset.

\begin{table}[t]
  \centering
  \caption{Markdown corpus summary.}
  \label{tab:data-collection-summary}
  \fontsize{8.15}{8.65}\selectfont
  \begin{tabular*}{\columnwidth}{@{\extracolsep{\fill}}lr@{}}
    \toprule
    Measure & Value \\
    \midrule
    Total Markdown files & 1,248 \\
    Source repositories & 276 \\
    Distinct commits ($\geq$1 Markdown file) & 7,446 \\
    Total Markdown file changes & 20,841 \\
    Commit-history span & 2025-07-31 to 2026-08-04 \\
    \bottomrule
  \end{tabular*}
\end{table}

\subsection{Instruction Taxonomy}
\label{sec:study-taxonomy}

We select a representative sample of 294 files from the 1,248 Markdown sources using Cochran's sample-size formula with finite-population correction, choosing 95\% confidence, a 5\% margin of error, and a conservative population proportion of \(p=0.5\).
We inspect each file's pinned latest snapshot, coding only its Markdown body
without expanding imports or inferring instructions from YAML frontmatter.
Coding is multi-label, as one file can contain several instruction types.

We use thematic analysis \cite{braun2012thematic}, organizing the process
into four steps \cite{chouchen2024chatgpt}:
\begin{enumerate}[leftmargin=*,itemsep=0.1em,topsep=0.25em]
  \item \textbf{Initial reading.} Two authors jointly review 30 files to
  familiarize themselves with the instructions and calibrate their coding.
  They then inspect the sampled bodies to identify what developers ask
  agents to do and the conditions placed on that work.
  \item \textbf{Generating descriptive labels.} The authors independently
  assign labels to instructions supported by the text. Labels describe
  the instruction's purpose rather than its wording. For example, limits
  on time, retries, or tokens fall under \textit{Resource Budget}. A file
  receives multiple labels when it contains different instruction types.
  \item \textbf{Reviewing and refining labels.} We compare related labels,
  review their scope against the bodies, and distinguish overlapping
  concepts. For instance, we merge \textit{Expert Role} and
  \textit{Role Scope} into \textit{Role Assignment}.
  \item \textbf{Grouping labels into themes.} We organize labels by their
  shared purpose and define the resulting categories. For example,
  \textit{Required Sequence}, \textit{Decision Branch}, and
  \textit{Within-Run Iteration} belong to \textit{Process}, as they specify
  how work proceeds. We report the taxonomy with definitions, frequencies,
  and examples from the files.
\end{enumerate}

\section{Results}

This section reports and discusses the results for our four research questions.

\subsection{RQ1: What are the characteristics of GitHub Agentic Workflows?}
\label{sec:rq1}

\noindent\rqpart{Motivation.} Writing a workflow in natural language does not necessarily make it easier to maintain compared to YAML. Before running, modifying, or copying a gh-aw workflow, developers need to review both its configuration and the instructions that guide the agent. Prior work on agent context files highlights readability and organization as concerns for developers who write and maintain these instructions \cite{chatlatanagulchai2026agentreadmes}. Our goal is to characterize the amount of text developers must review, its readability, and how it is organized, providing a baseline for creating and maintaining gh-aw workflows.

\noindent\rqpart{Approach.} To address RQ1, we analyze the latest available snapshots of the 1,248
Markdown files from three perspectives: size, structure, and readability.
We first count each file's body, frontmatter, and code-block lines. We then count heading levels per file and the most frequent used words in these headings to understand how developers divide these files into sections or phases. For the word cloud, we remove common English stopwords and standalone numbers to emphasize heading content, following prior word-cloud analyses \cite{johnson2025procedure,ali2022mapping}.

To measure readability, we use the Flesch Reading Ease (FRE) metric, where higher scores
indicate easier reading \cite{eleyan2020enhancing}, along with sentence and syllable counts. These measurements are applied on the complete Markdown body, including headings, tables, code, and template expressions.Finally, we summarize measurements using minimum, mean, median, and maximum values.

\noindent\rqpart{Results.} 

\begin{table}[t]
  \centering
  \caption{Markdown size, structure, and readability.}
  \label{tab:rq1-characteristics}
  \fontsize{8}{9.3}\selectfont
  \setlength{\tabcolsep}{2pt}
  \begin{tabular*}{\columnwidth}{@{\extracolsep{\fill}}lrrrr@{}}
    \toprule
    Metric & Min & Mean & Median & Max \\
    \midrule
    \multicolumn{5}{@{}l}{\normalsize\bfseries\itshape Size and structure} \\
    Body lines & 2 & 144 & 104 & 1,661 \\
    Frontmatter lines & 6 & 87 & 58 & 1,237 \\
    Code block lines & 0 & 31 & 5.5 & 603 \\
    \midrule
    \multicolumn{5}{@{}l}{\normalsize\bfseries\itshape Heading levels} \\
    Total headings (H1--H6) & 0 & 10.3 & 9 & 73 \\
    H1 & 0 & 0.9 & 1 & 19 \\
    H2 & 0 & 4.8 & 4 & 22 \\
    H3 & 0 & 4.0 & 3 & 44 \\
    H4 & 0 & 0.5 & 0 & 36 \\
    H5 & 0 & 0.0 & 0 & 3 \\
    \midrule
    \multicolumn{5}{@{}l}{\normalsize\bfseries\itshape Text and readability} \\
    Body words & 1 & 710 & 556.5 & 10,551 \\
    Sentences & 1 & 55 & 43 & 887 \\
    Syllables & 9 & 1,593 & 1,186 & 23,155 \\
    Pictographic symbols & 0 & 4.4 & 0 & 171 \\
    Flesch Reading Ease & -555.6 & 33.5 & 38.8 & 84.8 \\
    \bottomrule
  \end{tabular*}
\end{table}

\smallskip
\rqfinding{1}{gh-aw Markdown files contain a median of 9 headings per file,
with ``step,'' ``phase,'' and ``issue'' as the most frequent heading words.}

\begin{figure}[t]
  \centering
  \includegraphics[width=0.86\columnwidth]{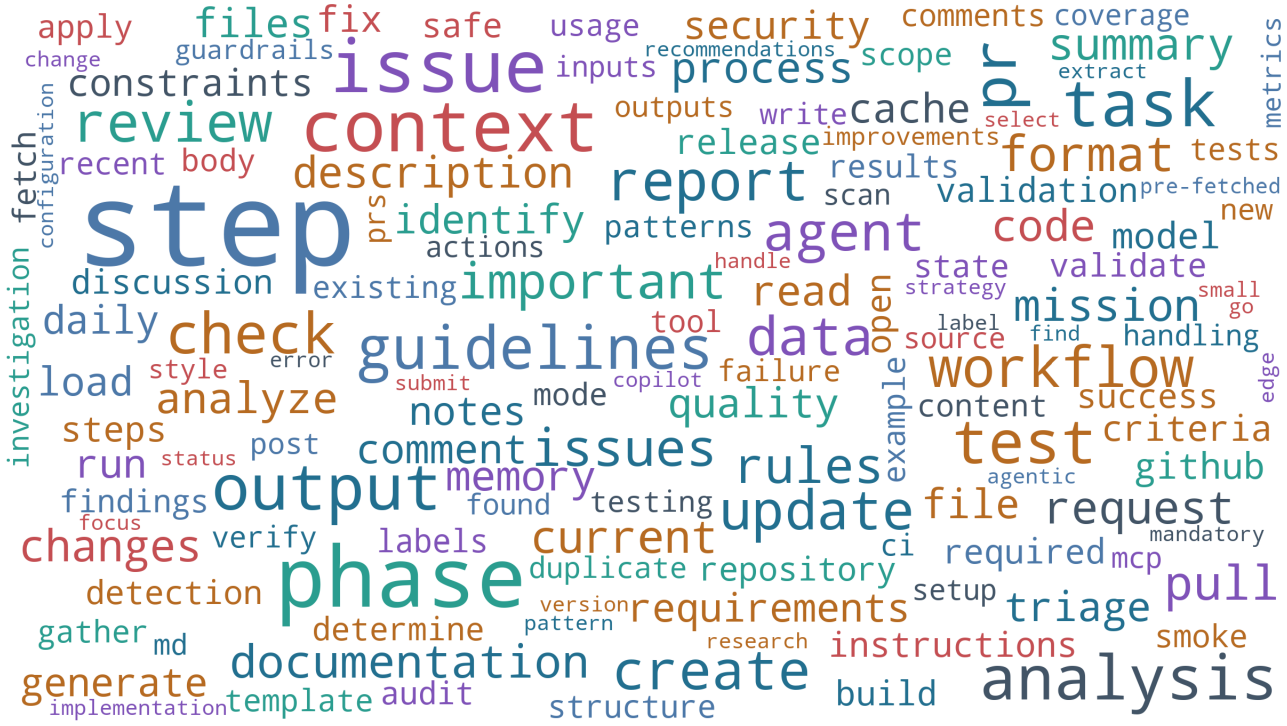}
  \caption{Frequent Markdown heading terms, sized by total occurrences.}
  \label{fig:rq1-heading-wordcloud}
\end{figure}

Table~\ref{tab:rq1-characteristics} summarizes workflow
size, structure, and readability, while Figure~\ref{fig:rq1-heading-wordcloud} shows the most frequent heading terms. In our set of markdown files, We find that heading counts range from zero to 73 per file
(Table~\ref{tab:rq1-characteristics}). Developers typically organize sections into a shallow hierarchy having H2 and H3 headings account for 85.5\% of all headings, with per file medians of four and three, respectively. 

At the file level, \textit{step}, \textit{issue}, and \textit{phase} appear in
headings in 31.3\%, 28.4\%, and 11.7\% of the files.
Upon further investigation of these terms, we noticed that \textit{step}
headings specify individual actions for the agent to follow, which aligns with the natural YAML workflow composition of steps. \textit{Issue} headings cover both handling
existing GitHub issues and reporting findings through new ones, as illustrated
by headings such as ``Issue Triage'' and ``Issue Format.'' Meanwhile,
\textit{phase} headings divide the work into numbered stages, such as gathering
context in Phase 1 and checking its completeness in Phase 2.
Other frequent terms include \textit{context}, \textit{task}, and \textit{output}, further emphasizing the nature of a workflow and context provided to the AI agents.

\smallskip
\rqfinding{2}{gh-aw Markdown files fall within the conventional difficult range to read, with a median Flesch Reading Ease score of 38.8.}

Our results suggest that gh-aw Markdown bodies are not easy to read under
the FRE scale, with a median of 38.8 in its conventional difficult range
(Table~\ref{tab:rq1-characteristics}). For comparison, prior studies report
mean scores of 29.5--41.1 for requirements/user stories
\cite{wouters2022crowd,bala2024chatgpt} and about 14--21 for SE abstracts
\cite{prechelt2025abstracts}, compared with medians of 56.9 for ChatGPT
responses and 68.8 for developer prompts \cite{ehsani2026chatgpt}.
These comparisons provide context rather than a direct ranking, since our
scores include the complete Markdown body rather than only natural-language text.

One possible explanation is that these bodies mix written instructions with
code blocks (62.1\% of files), template expressions (60.4\%), and pictographic
symbols (47.4\%). Code and template syntax can affect the word and sentence
counts used by FRE. Symbols also convey information beyond ordinary words.
For example, the smoke-proxy test workflow contains 171 symbols, using check
marks and crosses to report test outcomes instead of the English keywords
``success'' and ``failure'' \cite{github-gh-aw-mcpg-smoke-proxy-github-script}.
This mix may reflect a focus on agent execution rather than human reading,
but the scores alone do not establish that developers have little need to
read or review these files.

\subsection{RQ2: How are GitHub Agentic Workflows maintained over time?}
\label{sec:rq2}

\begin{figure*}[t]
  \centering
  \includegraphics[width=.93\textwidth]{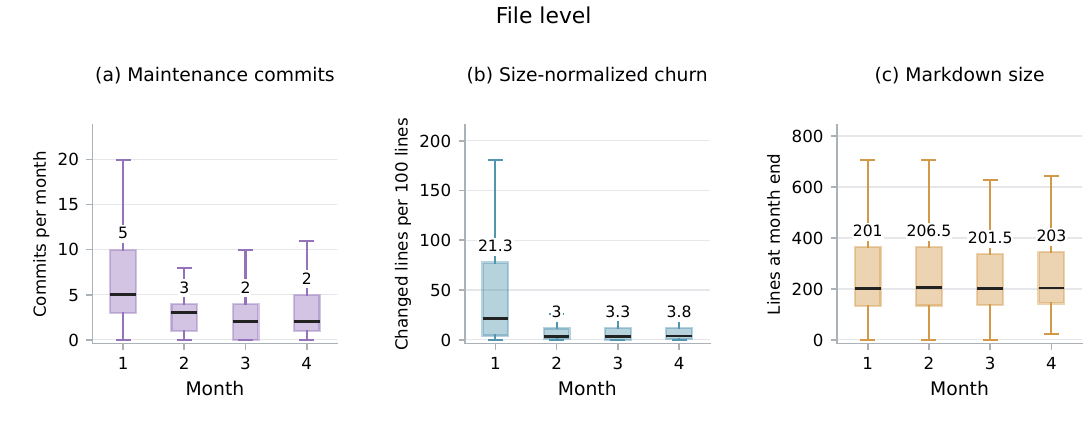}
  \caption{Monthly maintenance across 348 files. Commits and churn
  are monthly, not cumulative. Numbers show medians.}
  \label{fig:rq2-file-week}
\end{figure*}

\begin{figure*}[t]
  \centering
  \includegraphics[width=.93\textwidth]{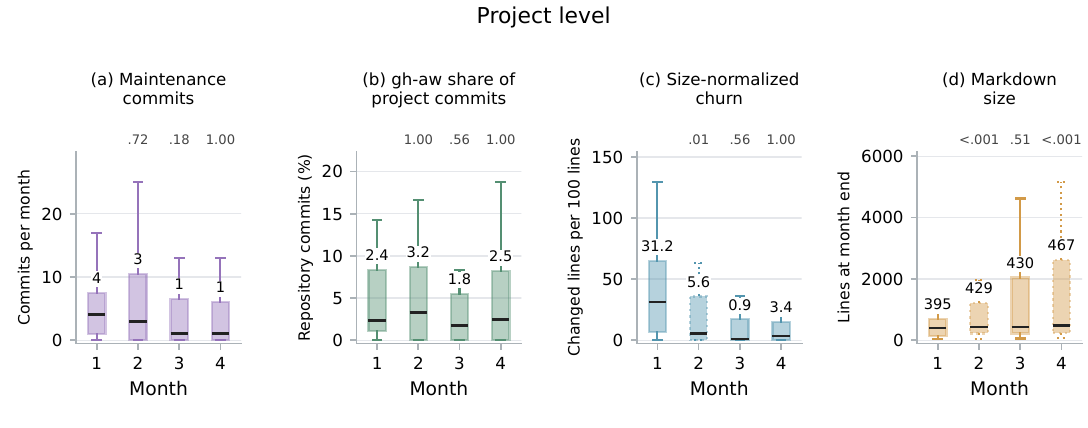}
  \caption{Monthly maintenance across 43 projects. Numbers show
  medians. Holm-adjusted \(p\)-values above the boxes compare each month
  with the previous month (significance at \(p<0.05\)).}
  \label{fig:rq2-project-week}
\end{figure*}

\begin{figure}[t]
  \centering
  \includegraphics[width=.95\columnwidth]{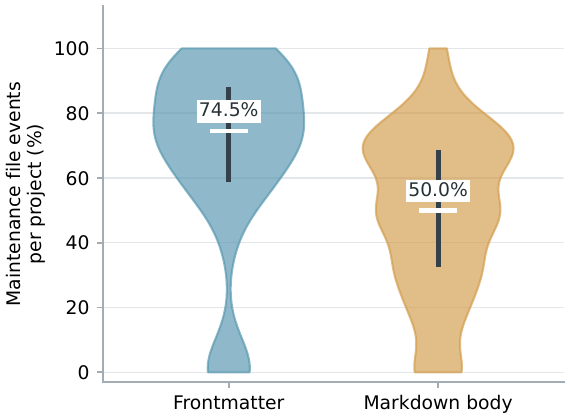}
  \caption{Percentage of each project's gh-aw Markdown maintenance changes
  touching frontmatter or body. Median values are shown.}
  \label{fig:rq2-project-location}
\end{figure}

\noindent\rqpart{Motivation.} In our dataset, 108 of 365 projects (29.6\%)
reference Markdown sources from another repository for at least one generated
workflow. This reuse makes the evolution of those sources relevant beyond their
original projects. Incorporating upstream changes and recompiling can alter the
generated YAML \cite{github-gh-aw-compilation}, potentially changing workflow
behavior in ways developers overlook without reviewing the updated source.
Rather than assume that reused sources remain fixed after creation, we examine
the pace and nature of their maintenance to understand the changes that
projects may inherit over time.

\noindent\rqpart{Approach.} To address RQ2, we analyze 348 gh-aw Markdown files from 43 repositories, each file having at least 120 days between its first and last
observed changes. Day 0 is the file's first observed appearance at its tracked path. For projects, day 0 is the earliest appearance of a collected gh-aw Markdown file. We track the first four months, including all collected Markdown files in each project and those introduced later.

We measure monthly commits, end-of-month size, and churn (added plus deleted
lines) per 100 lines, using the average Markdown size during the month
\cite{nagappan2005relative}. Initial file commits are excluded from maintenance,
and months without changes are retained. Projects sum file sizes and churn but
count each commit once. Inspired by prior comparisons of workflow maintenance
with project code \cite{valenzuela2024hidden}, we also divide maintenance commits by all repository commits in the same month. This contextualizes maintenance against overall project activity. We collect this baseline
through GitHub's commits API \cite{github-rest-commits}.

We compare each month with the preceding month using project-paired sign
tests \cite{nist2023signtest}, with Holm correction across the 12 comparisons
\cite{holm1979simple}. Months without repository commits have no defined
maintenance share. For change targets, each unique commit-Markdown-file pair is one event. Within each project, we divide events touching frontmatter or
body by all maintenance events, counting both-region events in both shares.

\noindent\rqpart{Results.} Figures~\ref{fig:rq2-file-week} and~\ref{fig:rq2-project-week} show maintenance activity and size across four months. Figure~\ref{fig:rq2-project-location} compares change targets across projects.

\smallskip
\rqfinding{3}{Gh-aw total Markdown size grows significantly during the first
4 months at the project level}

Figure~\ref{fig:rq2-project-week} shows that median total Markdown size per
project rises from 395 lines in month 1 to 429, 430, and 467 lines in
months 2, 3, and 4, respectively. The increases from month 1 to month 2
and from month 3 to month 4 are statistically significant (both
Holm-adjusted \(p<.001\)). The exception is month 2 to month 3
(adjusted \(p=.511\)). In contrast, median file size remains near
200 lines throughout the four months (Figure~\ref{fig:rq2-file-week}).
This contrast may reflect projects adding workflows rather than primarily
expanding existing ones. However, the file panel follows long-lived files,
whereas project totals also include newly introduced files, so these medians
alone do not establish the source of growth.

Moreover, median project-level churn drops from 31.2 to 5.6 changed lines
per 100 lines between months 1 and 2 (adjusted \(p=.011\)), with no
significant differences between later consecutive months. This suggests
more extensive revisions near adoption, followed by fewer edits relative
to workflow size as the project's collection grows. Nevertheless, the
share of project commits touching these workflows does not differ
significantly between consecutive months (all adjusted \(p\geq.561\),
Figure~\ref{fig:rq2-project-week}(b)). Thus, the early churn drop should
not be taken as evidence that workflow maintenance declines independently
of overall project activity.

\rqfinding{4}{At project level, a median of 74.5\% of Markdown maintenance
changes touch frontmatter, compared with 50.0\% touching the body.}

From Figure~\ref{fig:rq2-project-location}, we observe that the project-level medians suggest that maintenance more often touches frontmatter than body text during the observed period. Changes touching both regions enter
both percentages, and the three projects without maintenance changes receive
0\%. Across the 8,575 analyzed changes, counting each commit once per
Markdown file it touches, 44.8\% affect only frontmatter, 37.8\% only the
body, and 17.3\% both. The remaining 0.1\% have no content change or an
unknown location. These pooled percentages differ from the violin plot's
medians, which give each project equal weight.

Upon further investigation, frontmatter-only changes account for 36.7\%,
29.9\%, 54.5\%, and 71.7\% of all changes in months 1 through 4,
respectively, showing a larger share in the later months when changes are
pooled across projects. One possible explanation is that execution settings,
such as schedules, permissions, and tool access, may require adjustments
even when the workflow's underlying task remains unchanged. Frontmatter
maintenance may therefore reflect adapting how a workflow runs rather than
rewriting what its agent should do.

\subsection{RQ3: What agentic instructions do developers set in GitHub Agentic Workflows?}
\label{sec:rq3}

\noindent\rqpart{Motivation.} RQ1 characterizes the size, readability, and
structure of gh-aw Markdown files. However, these measurements do not reveal
what developers instruct agents to do or which rules they should follow.
The Markdown file defines the intended behavior and serves as the source
from which gh-aw generates the YAML workflow \cite{github-gh-aw-compilation}.
The generated \texttt{.lock.yml} files discourage direct edits with
the warning ``\textit{This file was automatically generated by gh-aw \ldots{}
DO NOT EDIT.}''
\cite{activiti-supply-chain-lock,awesome-copilot-duplicate-lock,pydantic-attention-triage-lock}.
RQ3 therefore qualitatively examines these instructions to build a taxonomy
and measure how often each type appears. The taxonomy brings together common
and less-used instructions that practitioners can consider when defining
their GitHub agentic workflows.

\noindent\rqpart{Approach.} To address RQ3, we apply the thematic analysis
described in Section~\ref{sec:study-taxonomy} to the sampled Markdown bodies.
We assess agreement over binary file-label decisions on shared labeled
files before resolution using percentage agreement and Cohen's kappa
\cite{cohen1960kappa}. Agreement was 83.61\% (\(\kappa=0.61\)), indicating
substantial agreement \cite{landis1977measurement}.
The two authors then met to reconcile disagreements, leading to a refined and fully agreed coding set. We calculate prevalence from the resolved labels, counting each label once per file and
excluding \textit{Cannot Label} cases.

\noindent\rqpart{Results.} 

\begin{table*}[h]
    \fontsize{8}{9}\selectfont
	\tabcolsep=0.2cm
  \centering
  \caption{Instruction taxonomy and prevalence after label resolution (288 files).}
  \label{tab:rq3-taxonomy-results}
  \renewcommand{\arraystretch}{1}
  \begin{tabular*}{\textwidth}{@{}>{\raggedright\arraybackslash}p{0.33\textwidth}>{\raggedright\arraybackslash}p{\dimexpr0.67\textwidth-2\tabcolsep\relax}@{}}
    \toprule
    Label and prevalence & Definition \\
    \hline
    \rowcolor{rq3category}\multicolumn{2}{@{}l@{}}{\bfseries TASK (97.6\%)} \\
    \hspace{0.45em}Evidence Investigation (51.7\%) & Investigates failures, researches topics, compares alternatives, or summarizes evidence. \\
    \hspace{0.45em}Artifact Change (33.3\%) & Creates, edits, repairs, translates, or removes code, documentation, or other project files. \\
    \hspace{0.45em}Artifact Review (25.0\%) & Assesses existing code, documents, or proposals and reports findings or improvements. \\
    \hspace{0.45em}Executable Validation (20.8\%) & Runs tests, builds, linters, benchmarks, or service and behavior checks. \\
    \hspace{0.45em}Work Item Management (17.4\%) & Triages, assigns, labels, closes, or updates issues and pull requests. \\
    \hline
    \rowcolor{rq3category}\multicolumn{2}{@{}l@{}}{\bfseries OUTPUT (95.8\%)} \\
    \hspace{0.45em}Control Signal (52.1\%) & Requires a status or control action, such as noop, labeling, closure, approval, or dispatch. \\
    \hspace{0.45em}New Thread (41.7\%) & Requires a new issue or discussion as the primary published result. \\
    \hspace{0.45em}Existing Thread Response (39.6\%) & Adds a comment, reply, or review to an existing issue, pull request, or discussion. \\
    \hspace{0.45em}Repository Change (29.9\%) & Requires a repository update through a commit, pull request, merge, or release change. \\
    \hspace{0.45em}Standalone Artifact (13.9\%) & Requires a standalone report, file, diagram, chart, archive, or upload. \\
    \hline
    \rowcolor{rq3category}\multicolumn{2}{@{}l@{}}{\bfseries CONSTRAINT (93.8\%)} \\
    \hspace{0.45em}Scope Boundary (76.7\%) & Restricts allowed files, repositories, actions, targets, or changes. \\
    \hspace{0.45em}Quality Standard (74.7\%) & Requires clear, accurate, concise, complete, or well-formatted work. \\
    \hspace{0.45em}Resource Budget (38.2\%) & Limits time, retries, tokens, cost, files inspected, issues created, or analysis depth. \\
    \hspace{0.45em}Evidence Grounding (36.1\%) & Requires facts, citations, or source locations to support claims and avoid speculation. \\
    \hline
    \rowcolor{rq3category}\multicolumn{2}{@{}l@{}}{\bfseries PROCESS (93.4\%)} \\
    \hspace{0.45em}Required Sequence (88.2\%) & Sets prerequisites or a required order of execution steps. \\
    \hspace{0.45em}Decision Branch (87.2\%) & Chooses actions by condition, including fallbacks, early stops, and error handling. \\
    \hspace{0.45em}Within-Run Iteration (35.8\%) & Repeats work within one run through loops, batches, retries, or review passes. \\
    \hspace{0.45em}Cross-Run Memory (18.1\%) & Reads or updates state retained between runs, such as past findings or processed items. \\
    \hspace{0.45em}Work Delegation (9.0\%) & Assigns a task to another agent, worker, skill, or workflow. \\
    \hline
    \rowcolor{rq3category}\multicolumn{2}{@{}l@{}}{\bfseries RESOURCE (89.9\%)} \\
    \hspace{0.45em}GitHub Interface (60.4\%) & Uses GitHub APIs, GitHub MCP tools, the gh CLI, or GitHub data retrieval. \\
    \hspace{0.45em}Local Workspace (57.6\%) & Uses local files, source code, logs, scripts, or general shell commands. \\
    \hspace{0.45em}Agent Resource (21.5\%) & Uses subagents, reusable skills, model calls, or agent memory services. \\
    \hspace{0.45em}Development Toolchain (18.8\%) & Uses tools for building, testing, linting, benchmarking, or static analysis. \\
    \hspace{0.45em}Web Retrieval (14.2\%) & Retrieves information from websites, search engines, feeds, docs, or external services. \\
    \hline
    \rowcolor{rq3category}\multicolumn{2}{@{}l@{}}{\bfseries CONTEXT (84.7\%)} \\
    \hspace{0.45em}Repository Context (70.1\%) & Describes repository structure or the relevant issue, pull request, branch, or event. \\
    \hspace{0.45em}Domain Context (34.0\%) & Explains subject knowledge, business rules, API conventions, or protocols. \\
    \hspace{0.45em}Execution Environment (34.0\%) & Describes available software, operating systems, containers, paths, or pre-fetched inputs. \\
    \hspace{0.45em}Historical Context (8.3\%) & Supplies information about earlier runs, releases, findings, baselines, or decisions. \\
    \hline
    \rowcolor{rq3category}\multicolumn{2}{@{}l@{}}{\bfseries EVALUATION (70.1\%)} \\
    \hspace{0.45em}Rule Conformance (47.2\%) & Checks an artifact against a specification, schema, style rule, or acceptance criterion. \\
    \hspace{0.45em}Risk Assessment (29.9\%) & Judges severity, likely impact, confidence, security risk, or priority. \\
    \hspace{0.45em}Execution Check (25.7\%) & Checks whether code builds, tests pass, services respond, or behavior matches expectations. \\
    \hspace{0.45em}Evidence Credibility (11.1\%) & Checks evidence reliability, recency, completeness, or agreement with other sources. \\
    \hline
    \rowcolor{rq3category}\multicolumn{2}{@{}l@{}}{\bfseries EXAMPLE (68.1\%)} \\
    \hspace{0.45em}Output Template (61.1\%) & Shows a report, message, table, or other output layout with fields to fill in. \\
    \hspace{0.45em}Technical Snippet (30.2\%) & Illustrates commands, code, queries, JSON, configuration, or diagram syntax. \\
    \hspace{0.45em}Worked Case (19.4\%) & Shows acceptable, unacceptable, or borderline cases to explain a rule or decision. \\
    \hline
    \rowcolor{rq3category}\multicolumn{2}{@{}l@{}}{\bfseries COMMUNICATION (63.9\%)} \\
    \hspace{0.45em}Role Assignment (46.9\%) & Assigns a specialist role, such as reviewer, research assistant, or language expert. \\
    \hspace{0.45em}Automation Disclosure (25.0\%) & Requires outputs to disclose automation, limitations, uncertainty, or missing checks. \\
    \hspace{0.45em}Audience Voice (20.1\%) & Specifies the audience, language, tone, or communication style. \\
    \hline
    \rowcolor{rq3category}\multicolumn{2}{@{}l@{}}{\bfseries SAFETY (25.0\%)} \\
    \hspace{0.45em}Execution Isolation (13.9\%) & Specifies sandbox, filesystem, network, permission, or read-only restrictions. \\
    \hspace{0.45em}Prompt Injection Defense (9.4\%) & Addresses untrusted instructions and which sources may direct the agent. \\
    \hspace{0.45em}Information Confidentiality (7.3\%) & Addresses secrets, credentials, personal data, or other sensitive information. \\
    \hspace{0.45em}Vulnerability Handling (6.3\%) & Directs finding, assessing, fixing, or reporting exploitable weaknesses. \\
    \bottomrule
    \multicolumn{2}{@{}p{\textwidth}@{}}{\fontsize{8}{9}\selectfont Denominator: 288 files with resolved semantic labels. Six Cannot Label cases are excluded from the corrected 294-file sample. Each label counts once per file. Multi-label percentages do not sum to 100\%.}
  \end{tabular*}
\end{table*}

\smallskip
\rqfinding{5}{Tasks, outputs, constraints, and execution processes are common
elements of gh-aw Markdown, each appearing in at least 93.4\% of labeled files.}
\nopagebreak[4]

Table~\ref{tab:rq3-taxonomy-results} presents the
instruction taxonomy and its prevalence, while
Snippets~\ref{snip:rq3-task-evidence-investigation} to
\ref{snip:rq3-safety-prompt-injection-defense} illustrate these instructions in gh-aw Markdown files. The dominant categories show that workflow Markdown files usually combine a
substantive task with explicit delivery, behavioral, and execution-control
instructions. Most coded files contain TASK (97.6\%), OUTPUT (95.8\%),
CONSTRAINT (93.8\%), and PROCESS (93.4\%) labels. At the subcategory level, the
most frequent patterns are \rqsubcat{Required Sequence} (88.2\%),
\rqsubcat{Decision Branch} (87.2\%), \rqsubcat{Scope Boundary} (76.7\%),
\rqsubcat{Quality Standard} (74.7\%), and \rqsubcat{Repository Context} (70.1\%).
Categories and their subcategories are presented in descending prevalence order
in both the table and the descriptions below, with ties ordered alphabetically.
All descriptions use the same file-level denominator. The two most frequent
subcategories in each category have boxed excerpts, while the remaining examples
appear inline. Excerpts preserve the source's Markdown formatting, with
\texttt{[...]} marking internal omissions. Repository and filename labels link
to the pinned sources, as do the compact source links after inline quotations.
Every example's source carries its illustrated label
in the resolved export.

\medskip
\noindent\faIcon{tasks}~\textbf{1 TASK (97.6\%).} Task instructions define the
work the agent should perform. \rqsubcat{Evidence Investigation} (51.7\%)
includes diagnosing failures, researching a topic, comparing alternatives, and
summarizing evidence. For example, in Snippet~\ref{snip:rq3-task-evidence-investigation},
OpenVINO's CI doctor investigates every failing pipeline on a pull request
and identifies possible remedies. The agent must explain the failures rather
than simply report that the checks failed.

\begin{rq3mdexample}{Evidence Investigation}{openvinotoolkit/openvino}{ci-doctor.md}{https://github.com/openvinotoolkit/openvino/blob/04972c078a3b24ed9ac9e29491b1f47f9c6f632d/.github/workflows/ci-doctor.md\#L49-L49}{snip:rq3-task-evidence-investigation}
Your mission is to investigate every failing pipeline on that PR and post a single, consolidated summary comment describing each failure and a possible remedy.\par
\end{rq3mdexample}

\noindent \rqsubcat{Artifact Change} (33.3\%) covers creating, editing,
repairing, translating, or removing code, documentation, tests, and other project
files. For instance, in Snippet~\ref{snip:rq3-task-artifact-change}, OpenMS's
workflow adds missing changelog entries for recent changes. It asks the agent to update the
file, not merely report that entries are missing.

\begin{rq3mdexample}{Artifact Change}{OpenMS/OpenMS}{changelog-sync.md}{https://github.com/OpenMS/OpenMS/blob/a4e55651230626206f7c89203fccc05b368e18f5/.github/workflows/changelog-sync.md\#L31-L33}{snip:rq3-task-artifact-change}
\rqmdheading{2}{Your Task}
Your goal is to identify noteworthy recent code changes (from the last 7 days) that are not yet reflected in the top-level \rqmdinline{CHANGELOG} file and open a pull request with the necessary updates.\par
\end{rq3mdexample}

\noindent \rqsubcat{Artifact Review} (25.0\%) asks for an assessment of existing
code, documentation, specifications, or proposals, with findings or suggested
improvements. For example, a specification planner asks the agent to
\textit{``review repository specification documents''}, making the existing
specifications the subject of its assessment.\rqexamplesource{https://github.com/github/gh-aw/blob/8fcdd9f048c3f75a5046372f4000abef92cdac1b/.github/workflows/daily-spdd-spec-planner.md\#L96}
\rqsubcat{Executable Validation} (20.8\%) asks the agent to run tests,
builds, linters, benchmarks, or service checks. For instance, one workflow asks
the agent to \textit{``Test GitHub
MCP tool access under this guard policy by performing these operations and
reporting results.''}\rqexamplesource{https://github.com/github/gh-aw/blob/cd0b423cc66f05fbe79faac926c3358274f9c0be/.github/workflows/smoke-agent-public-none.md\#L60}
\rqsubcat{Work Item Management}
(17.4\%) covers triaging, assigning, labeling, closing, or updating issues and
pull requests. For example, Handbookbot requests \textit{``Process unlabelled
issues and PRs. Apply labels from: [...] Remove misapplied labels.''}, covering
both assigning and correcting labels.\rqexamplesource{https://github.com/NikiforovAll/claude-code-rules/blob/79fbbd2187080066e54ac81cc22fb199dd81e1fc/.github/workflows/handbookbot.md\#L189}

\medskip
\noindent\faIcon{file-alt}~\textbf{2 OUTPUT (95.8\%).} Output instructions
specify how and where the agent delivers its result. \rqsubcat{Control Signal}
(52.1\%) covers status or control actions, such as returning \texttt{noop},
applying a label, closing an issue, or approving a pull request. For example,
in Snippet~\ref{snip:rq3-output-control-signal}, Thaw's triage workflow requires a
priority label to be attached to the issue. The result is recorded in the
issue's labels rather than only described in a response.

\begin{rq3mdexample}{Control Signal}{thaw-app/Thaw}{issue-triage.md}{https://github.com/thaw-app/Thaw/blob/2168eb56fb19f5fea69103991646b4021dd1b11b/.github/workflows/issue-triage.md\#L124-L126}{snip:rq3-output-control-signal}
\rqmdheading{3}{3. Assign a Priority Label}
For \textbf{bug} and \textbf{regression} issues, assess severity and impact, then apply \textbf{exactly one} priority label using \rqmdinline{add\_labels}:\par
\end{rq3mdexample}

\noindent \rqsubcat{New Thread} (41.7\%) requires publishing the result as a
new issue or discussion, such as a bug report, investigation, or work plan.
For instance, in Snippet~\ref{snip:rq3-output-new-thread}, Elastic's CI detective
opens a separate issue for a newly identified failure, giving that failure its
own discussion thread.

\begin{rq3mdexample}{New Thread}{elastic/ai-github-actions}{gh-aw-branch-actions-detective.md}{https://github.com/elastic/ai-github-actions/blob/038d4f7b0986ad158aa9efe8f28a41df4e71b48f/.github/workflows/gh-aw-branch-actions-detective.md\#L141-L144}{snip:rq3-output-new-thread}
\textbf{If this is a new or distinct failure:} Call \rqmdinline{create\_issue} with the following format:\par
\textbf{Issue title:} Brief summary of the CI failure\par
\end{rq3mdexample}

\noindent \rqsubcat{Existing Thread Response} (39.6\%) covers comments,
replies, and reviews attached to an existing issue, pull request, or discussion.
For example, Issue Monster instructs the agent to \textit{``For each issue you
assign, use the [...] tool [...] to add a comment''}, keeping the response in
the existing conversation.\rqexamplesource{https://github.com/github/gh-aw-firewall/blob/02cb22631932cfbeae4d633edf8f0202dbbf6b5d/.github/workflows/issue-monster.md\#L391-L393}

\noindent \rqsubcat{Repository Change} (29.9\%) requires a commit, pull
request, merge, or release update. For example, the documentation updater in
\texttt{microsoft/agentrc} specifies \textit{``Call the
\texttt{create\_pull\_request} MCP tool''}, requiring the documentation edits
to be submitted as a pull request.\rqexamplesource{https://github.com/microsoft/agentrc/blob/a9a15ad42ab2e7896971fea09a8de2a34d5ca710/.github/workflows/daily-doc-updater.md\#L168-L179}
\rqsubcat{Standalone Artifact} (13.9\%) asks
for a separate report, file, chart, diagram, or archive. For instance, a smoke
test requests \textit{``Write a test results file to [...] with a brief summary.''},
giving the run a separate results file.\rqexamplesource{https://github.com/github/gh-aw-mcpg/blob/45a6fb81de50cc5e4c7dd0232522492a74ca8160/.github/workflows/smoke-allowonly.md\#L139}

\medskip
\noindent\faIcon{exclamation-triangle}~\textbf{3 CONSTRAINT (93.8\%).}
Constraint instructions limit the agent's actions and set standards for its
work. \rqsubcat{Scope Boundary} (76.7\%) specifies allowed files, repositories,
actions, or types of changes. For example, in
Snippet~\ref{snip:rq3-constraint-scope-boundary}, Elastic's issue-response workflow allows
local modifications but prohibits direct commits or pushes to the repository.
This separates the changes the agent may prepare from the repository actions
it may perform.

\begin{rq3mdexample}{Scope Boundary}{elastic/ai-github-actions}{gh-aw-mention-in-issue.md}{https://github.com/elastic/ai-github-actions/blob/038d4f7b0986ad158aa9efe8f28a41df4e71b48f/.github/workflows/gh-aw-mention-in-issue.md\#L108-L111}{snip:rq3-constraint-scope-boundary}
\rqmdheading{2}{Constraints}
\begin{itemize}[leftmargin=1.3em,itemsep=1pt,parsep=0pt,topsep=2pt]
\item \textbf{CAN}: Read files, search code, modify files locally, run tests and commands, comment on issues, create pull requests, create issues
\item \textbf{CANNOT}: Directly push or commit to the repository --- use \rqmdinline{ready\_to\_make\_pr} then \rqmdinline{create\_pull\_request} to propose changes
\end{itemize}
\end{rq3mdexample}

\noindent \rqsubcat{Quality Standard} (74.7\%) sets expectations for clear,
accurate, complete, well-formatted, or maintainable work. For instance, in
Snippet~\ref{snip:rq3-constraint-quality-standard}, the code-simplifier
requires readable changes that preserve behavior and follow project conventions.
Shorter code alone does
not meet its stated standard.

\begin{rq3mdexample}{Quality Standard}{githubnext/agentics}{code-simplifier.md}{https://github.com/githubnext/agentics/blob/899b885494293202a03b67147ebf4ff419d08498/workflows/code-simplifier.md\#L286-L290}{snip:rq3-constraint-quality-standard}
\rqmdheading{3}{Quality Standards}
\begin{itemize}[leftmargin=1.3em,itemsep=1pt,parsep=0pt,topsep=2pt]
\item \textbf{Test first}: Always run tests after simplifications (when available)
\item \textbf{Preserve behavior}: Functionality must remain identical
\item \textbf{Follow conventions}: Apply project-specific patterns consistently
\item \textbf{Clear over clever}: Prioritize readability and maintainability
\end{itemize}
\end{rq3mdexample}

\noindent \rqsubcat{Resource Budget} (38.2\%) limits time, retries, tokens,
cost, files inspected, issues created, or analysis depth. For example,
\texttt{news-weekly-review.md} instructs \textit{``Plan to call [...] by agent
minute 42 (hard deadline 45)''}, leaving a three-minute buffer for
publication.\rqexamplesource{https://github.com/Hack23/riksdagsmonitor/blob/18461d77f83dbce1774875fa4683773d5664bdb6/.github/workflows/news-weekly-review.md\#L310}
\rqsubcat{Evidence Grounding} (36.1\%) requires supporting facts, citations,
or source locations and prohibits invented claims. For instance, one workflow states:
\textit{``Be factual and specific. Cite issue numbers.''}, tying claims to
traceable issue records.\rqexamplesource{https://github.com/chrizbo/agentics-beyond-code/blob/f2f40e1eb40456327ac1afd92f6a54229d6d278e/.github/workflows/launch-readiness.md\#L336}

\medskip
\noindent\faIcon{project-diagram}~\textbf{4 PROCESS (93.4\%).} Process
instructions specify the order, repetition, conditions, and division of work.
\rqsubcat{Required Sequence} (88.2\%) sets prerequisites or an ordered series
of steps, such as checking CI before editing and testing before submitting.
For example, in Snippet~\ref{snip:rq3-process-required-sequence}, the avenger
workflow checks CI status before updating its checkout, making the order explicit.

\begin{rq3mdexample}{Required Sequence}{github/gh-aw}{avenger.md}{https://github.com/github/gh-aw/blob/9c4703514b8f99867ae28dd8cb234faa5da91138/.github/workflows/avenger.md\#L144-L158}{snip:rq3-process-required-sequence}
\rqmdheading{2}{Step 0: Verify CI Status}
Before doing anything:\par
[...]\par
\rqmdheading{2}{Step 1: Merge origin/main}
Bring your checkout up to date with the latest main branch:\par
\end{rq3mdexample}

\noindent \rqsubcat{Decision Branch} (87.2\%) specifies what to do when a
condition holds, including fallback actions, early stops, and error handling.
For instance, in Snippet~\ref{snip:rq3-process-decision-branch}, the documentation
validator inspects logs and opens an issue if deployment fails, but continues
with site validation if it succeeds.

\begin{rq3mdexample}{Decision Branch}{tosin2013/mcp-adr-analysis-server}{docs-site-validator.md}{https://github.com/tosin2013/mcp-adr-analysis-server/blob/8f650881f6db2dbe131f4472da3192de4eec7a3f/.github/workflows/docs-site-validator.md\#L45-L47}{snip:rq3-process-decision-branch}
\rqmdheading{2}{When triggered by workflow\_run}
If the Deploy Docusaurus workflow \textbf{failed}, check the deployment logs and create an issue about the deployment failure. If it \textbf{succeeded}, proceed with the full validation below.\par
\end{rq3mdexample}

\noindent \rqsubcat{Within-Run Iteration} (35.8\%) repeats work within one run,
such as processing each file, retrying a failed command, or making another
review pass. For example, a performance scanner instructs \textit{``For each
pattern below, run the bash commands, then read the flagged files''}, repeating
the inspection for each pattern.\rqexamplesource{https://github.com/microsoft/apm/blob/ee1f586b5f7f82e56d54fe47d4e5110ab162dc1a/.github/workflows/perf-scan.md\#L74-L75}
\rqsubcat{Cross-Run Memory} (18.1\%) reads or
updates state retained between runs, such as processed-issue lists, previous
findings, or cleanup notes. For instance, one workflow asks the agent to \textit{``check the
cache folder for notes about previous cleanups''}, carrying information from
earlier runs into the current cleanup.\rqexamplesource{https://github.com/github/gh-aw/blob/9c4703514b8f99867ae28dd8cb234faa5da91138/.github/workflows/unbloat-docs.md\#L227-L232}

\noindent \rqsubcat{Work Delegation} (9.0\%) assigns a task to another agent,
worker, skill, or workflow. For example, Elastic's breaking-change detector
requests \textit{``spawn 3 \texttt{general-purpose} sub-agents, each analyzing
the recent commits from a different angle''}. It divides the analysis among
agents that examine interface changes, documented guarantees, and downstream
compatibility.\rqexamplesource{https://github.com/elastic/ai-github-actions/blob/038d4f7b0986ad158aa9efe8f28a41df4e71b48f/.github/workflows/gh-aw-breaking-change-detector.md\#L114}

\medskip
\noindent\faIcon{tools}~\textbf{5 RESOURCE (89.9\%).} Resource instructions
identify tools and information sources the agent should use.
\rqsubcat{GitHub Interface} (60.4\%) includes GitHub APIs, GitHub MCP tools,
the \texttt{gh} CLI, and retrieval of issues, pull requests, or workflow runs.
For example, in Snippet~\ref{snip:rq3-resource-github-interface}, the OpenCode
smoke test directs the agent to retrieve pull-request details through GitHub
MCP tools, explicitly naming the interface it should use to access repository
information.

\begin{rq3mdexample}{GitHub Interface}{github/gh-aw}{smoke-opencode.md}{https://github.com/github/gh-aw/blob/87b330ed76a605c67633931d9fa321a674fcf80e/.github/workflows/smoke-opencode.md\#L74-L74}{snip:rq3-resource-github-interface}
\begin{enumerate}[label=\arabic*.,leftmargin=1.5em,itemsep=1pt,parsep=0pt,topsep=2pt,start=1]
\item \textbf{GitHub MCP Testing}: Use GitHub MCP tools to fetch details of exactly 2 merged pull requests from \$\{\{ github.repository \}\} (title and number only)
\end{enumerate}
\end{rq3mdexample}

\noindent \rqsubcat{Local Workspace} (57.6\%) directs the agent to read or
manipulate local files, source code, logs, and scripts, including shell commands
such as \texttt{find}, \texttt{cat}, and \texttt{git diff}. For instance, in
Snippet~\ref{snip:rq3-resource-local-workspace}, the SDK consistency reviewer
explicitly asks the agent to read files in other SDK directories. Those local
implementations provide the material for its cross-language comparison.
Before resolution, this label had
the lowest agreement (61.81\%, \(\kappa=0.238\)). Its assignment to 145 files
by the primary annotator versus 81 by the second suggests different treatment
of local inspection and shell commands within broader GitHub tasks.

\begin{rq3mdexample}{Local Workspace}{github/copilot-sdk}{sdk-consistency-review.md}{https://github.com/github/copilot-sdk/blob/f36e6bed37ca35ea278e9b5f1ff6af5d56c7566d/.github/workflows/sdk-consistency-review.md\#L91-L93}{snip:rq3-resource-local-workspace}
\begin{enumerate}[label=\arabic*.,leftmargin=1.5em,itemsep=1pt,parsep=0pt,topsep=2pt,start=4]
\item \textbf{Cross-reference other SDKs}: Check if the equivalent functionality exists in other language implementations:
\begin{itemize}[leftmargin=1.3em,itemsep=1pt,parsep=0pt,topsep=2pt]
\item Read the corresponding files in other SDK directories
\item Compare method signatures, behavior, and documentation
\end{itemize}
\end{enumerate}
\end{rq3mdexample}

\noindent \rqsubcat{Agent Resource} (21.5\%) directs use of subagents, reusable
skills, model calls, or agent memory services. For example,
\texttt{journey-e2e-test.md} directs the agent to read a reusable skill before
running a selected journey: \textit{``Read
\nolinkurl{.github/skills/journey-runner/SKILL.md} to understand the execution
pipeline.''} The workflow points to a reusable skill for guidance on carrying out that
work.\rqexamplesource{https://github.com/DanWahlin/github-azure-agentic-journeys/blob/e86f7211caa5400cdbfb533ce777d1745c1c2a2e/.github/workflows/journey-e2e-test.md\#L191-L197}

\noindent \rqsubcat{Development Toolchain} (18.8\%) names tools for building,
testing, linting, benchmarking, or static analysis, such as Gradle, Clang, and
Playwright. For instance, one smoke test requests \textit{``Use the playwright
tools to navigate to https://github.com''}, naming the browser tools for its
page-title check.\rqexamplesource{https://github.com/github/gh-aw-firewall/blob/c2e79f2cd0637f016684acb926c420f007c02acd/.github/workflows/smoke-codex.md\#L127}
\rqsubcat{Web Retrieval} (14.2\%) requests information from websites,
search engines, feeds, documentation, or external services. For example, a
content-campaign workflow asks the agent to \textit{``confirm whether content
is relevant to the intent [...] use the Tavily search tool''}, using web
search to check whether technologies mentioned in the content are
outdated.\rqexamplesource{https://github.com/SSWConsulting/SSW.Rules.Content/blob/0a9e7b77dc698a3fb4ea446221daf3c33523439b/.github/workflows/content-campaign.md\#L209}

\medskip
\noindent\faIcon{info-circle}~\textbf{6 CONTEXT (84.7\%).} Context provides
background for interpreting the task. \rqsubcat{Repository Context} (70.1\%)
supplies repository structure or details of the triggering issue, pull request,
branch, release, or commit. For example, in
Snippet~\ref{snip:rq3-context-repository-context}, Handbookbot identifies its
repository as a plugin monorepo with a separate documentation website. It explains what the
directories contain, without itself asking the agent to change them.

\begin{rq3mdexample}{Repository Context}{NikiforovAll/claude-code-rules}{handbookbot.md}{https://github.com/NikiforovAll/claude-code-rules/blob/79fbbd2187080066e54ac81cc22fb199dd81e1fc/.github/workflows/handbookbot.md\#L161-L165}{snip:rq3-context-repository-context}
\rqmdheading{2}{Context}
This is a Claude Code plugin monorepo containing:\par
\begin{itemize}[leftmargin=1.3em,itemsep=1pt,parsep=0pt,topsep=2pt]
\item Multiple distributable plugins in \rqmdinline{plugins/}
\item A Docusaurus documentation website in \rqmdinline{website/}
\end{itemize}
\end{rq3mdexample}

\noindent \rqsubcat{Domain Context} (34.0\%) supplies subject knowledge, such
as API conventions, compatibility rules, business concepts, or protocol
definitions. For instance, in Snippet~\ref{snip:rq3-context-domain-context}, the
model-mapping updater lists which API endpoints it expects different model families to support.
These rules guide its mapping task. They describe assumptions in the collected
snapshot, not a claim about current model support.

\begin{rq3mdexample}{ Domain Context}{github/gh-aw-firewall}{model-api-mapping-updater.md}{https://github.com/github/gh-aw-firewall/blob/5e5ae7bf4013bc439776e5a4e819908a076e4d3e/.github/workflows/model-api-mapping-updater.md\#L47-L50}{snip:rq3-context-domain-context}
\begin{enumerate}[label=\arabic*.,leftmargin=1.5em,itemsep=1pt,parsep=0pt,topsep=2pt,start=3]
\item For OpenAI models:
\begin{itemize}[leftmargin=1.3em,itemsep=1pt,parsep=0pt,topsep=2pt]
\item GPT-5.x family and newer \(\rightarrow\) \rqmdinline{responses} only (unless docs explicitly state chat/completions support)
\item o-series reasoning models (o1, o3, o4) \(\rightarrow\) check docs for dual support
\item GPT-4.x and older \(\rightarrow\) typically \rqmdinline{chat\_completions} (some support both)
\end{itemize}
\end{enumerate}
\end{rq3mdexample}

\noindent \rqsubcat{Execution Environment} (34.0\%) describes available
software, operating systems, containers, paths, or pre-fetched inputs. For instance, one
workflow explains that \textit{``The service containers run on the host, and
AWF routes traffic through the host gateway.''} This describes the network
environment in which the agent operates.\rqexamplesource{https://github.com/github/gh-aw/blob/1ac664c82b697ecf0c9a84222f48bbd8fd11e93a/.github/workflows/smoke-service-ports.md\#L61}
\rqsubcat{Historical Context} (8.3\%) supplies information about earlier runs,
releases, findings, or decisions. For example,
\texttt{flaky-test-catcher.md} in Elastic's Terraform provider repository
states that \textit{``A deterministic pre-activation step has already queried
CI run history and computed issue capacity for this run.''} It supplies
failed-run identifiers and counts as context for the agent's
investigation.\rqexamplesource{https://github.com/elastic/terraform-provider-elasticstack/blob/bc6470822986b0916f7eb0a2361046924f62ec5d/.github/workflows/flaky-test-catcher.md\#L77-L85}

\medskip
\noindent\faIcon{check-circle}~\textbf{7 EVALUATION (70.1\%).} Evaluation
instructions specify what the agent must check and how to judge the result.
\rqsubcat{Rule Conformance} (47.2\%) checks code, documentation, or other
artifacts against a specification, style rule, schema, or acceptance criterion.
For example, in Snippet~\ref{snip:rq3-evaluation-rule-conformance}, the MCP checker
verifies response structure and checks that error responses set \texttt{isError}.
It checks compliance with a protocol, not just whether a tool runs.

\begin{rq3mdexample}{Rule Conformance}{tosin2013/mcp-adr-analysis-server}{mcp-tool-checker.md}{https://github.com/tosin2013/mcp-adr-analysis-server/blob/72f88fa814801e4acb65e1eaf7f4f0437de3fdef/.github/workflows/mcp-tool-checker.md\#L62-L70}{snip:rq3-evaluation-rule-conformance}
\rqmdheading{3}{Check 1: Tool Response Pattern Consistency}
[...]\par
\begin{enumerate}[label=\arabic*.,leftmargin=1.5em,itemsep=1pt,parsep=0pt,topsep=2pt,start=1]
\item Read the file
\item Find all \rqmdinline{return} statements that produce tool responses
\item Verify each returns \rqmdinline{\{ content: [\{ type: "text", text: ... \}] \}}
\item Flag any tool that returns a different structure
\item Check that error responses set \rqmdinline{isError: true}
\end{enumerate}
\end{rq3mdexample}

\noindent \rqsubcat{Risk Assessment} (29.9\%) judges severity, likely impact,
confidence, security risk, or priority. For instance, in
Snippet~\ref{snip:rq3-evaluation-risk-assessment}, the change-risk assessor classifies
changes as breaking, risky, or safe according to their effect on downstream users. This asks the agent
to assess consequences, not merely list changed files.

\begin{rq3mdexample}{Risk Assessment}{githubnext/ado-aw}{risk.md}{https://github.com/githubnext/ado-aw/blob/603279af578e112f6987e2bbc50a6322e93eeab6/.github/workflows/risk.md\#L79-L79}{snip:rq3-evaluation-risk-assessment}
\begin{enumerate}[label=\arabic*.,leftmargin=1.5em,itemsep=1pt,parsep=0pt,topsep=2pt,start=4]
\item For each risk found, assess severity: \textbf{breaking} (consumers will fail), \textbf{risky} (may cause subtle issues), or \textbf{safe} (backward-compatible)
\end{enumerate}
\end{rq3mdexample}

\noindent \rqsubcat{Execution Check} (25.7\%) evaluates whether code builds,
tests pass, a service responds, or observed behavior matches expectations.
For example, the code-simplifier instructs \textit{``Run tests before and after
to ensure no behavioral changes''}, using the test outcomes to check that
the edits preserve behavior.\rqexamplesource{https://github.com/githubnext/agentics/blob/899b885494293202a03b67147ebf4ff419d08498/workflows/code-simplifier.md\#L110}

\noindent \rqsubcat{Evidence Credibility} (11.1\%) checks whether evidence is
reliable, current, complete, or supported by other sources. For instance, the Scout workflow
asks the agent to evaluate \textit{``Authority: Source credibility and
expertise''} and \textit{``Recency: How current the information is''}.
It also requests \textit{``Cross-reference information from multiple sources''},
checking the basis of the evidence rather than accepting a search result at
face value.\rqexamplesource{https://github.com/github/gh-aw/blob/1fafcdbb18ce775b11190e86300e5843be5bf92c/.github/workflows/scout.md\#L121-L128}

\medskip
\noindent\faIcon{code}~\textbf{8 EXAMPLE (68.1\%).} Examples show concrete
formats, code, or cases that clarify the instructions. \rqsubcat{Output
Template} (61.1\%) provides a report layout, message skeleton, table, or other
structure with fields to fill in. For example, in
Snippet~\ref{snip:rq3-example-output-template}, the Ollama smoke test supplies an
issue-title pattern and a Haiku section with a placeholder. The example shows where the
generated content belongs.

\begin{rq3mdexample}{Output Template}{github/gh-aw}{daily-byok-ollama-test.md}{https://github.com/github/gh-aw/blob/c2d74ef75dd7998888d7266898529716afa16f51/.github/workflows/daily-byok-ollama-test.md\#L115-L128}{snip:rq3-example-output-template}
Then create an issue with:\par
\begin{itemize}[leftmargin=1.3em,itemsep=1pt,parsep=0pt,topsep=2pt]
\item Title: \rqmdinline{BYOK Ollama Test --- \$\{\{ github.run\_id \}\}}
\item Body:
\begin{lstlisting}[style=rq3markdown]
[...]
### Haiku

<your haiku here>
\end{lstlisting}
\end{itemize}
\end{rq3mdexample}

\noindent \rqsubcat{Technical Snippet} (30.2\%) provides sample commands, code,
queries, JSON, configuration, or diagram syntax. For instance, in
Snippet~\ref{snip:rq3-example-technical-snippet}, the integrity-filtering audit in
\texttt{github/gh-aw-mcpg} gives an explicitly marked example command for counting
integrity events in logs. It illustrates a possible
query, rather than merely naming a tool or prescribing a mandatory command.

\begin{rq3mdexample}{Technical Snippet}{github/gh-aw-mcpg}{integrity-filtering-audit.md}{https://github.com/github/gh-aw-mcpg/blob/45a6fb81de50cc5e4c7dd0232522492a74ca8160/.github/workflows/integrity-filtering-audit.md\#L131-L133}{snip:rq3-example-technical-snippet}
\begin{lstlisting}[style=rq3markdown]
# Example: Count DIFC events in JSONL
grep -c 'difc_integrity' "$ARTIFACT_DIR"/*/mcp-logs/rpc-messages.jsonl 2>/dev/null || true
\end{lstlisting}
\end{rq3mdexample}

\noindent \rqsubcat{Worked Case} examples (19.4\%) illustrate acceptable,
unacceptable, or borderline cases, such as which changes to flag or which label
to choose. For example, a Go API reviewer accepts language-specific API forms
when equivalent functionality exists in other SDKs: \textit{``do not flag it
solely because the shape is idiomatic to Go.''} Its contrasting
\textit{``Bad: Divergent behavior''} case concerns changes that alter behavior
relative to the .NET and Python implementations.\rqexamplesource{https://github.com/microsoft/agent-framework-go/blob/a55fb3ec877409f92b3c0d1cffd0504ab95af354/.github/workflows/go-api-consistency-review.md\#L170-L176}

\medskip
\noindent\faIcon{comments}~\textbf{9 COMMUNICATION (63.9\%).} Communication
instructions shape how the agent presents itself and addresses readers.
\rqsubcat{Role Assignment} (46.9\%) assigns a role such as SDK reviewer, research
assistant, or language expert. For example, in
Snippet~\ref{snip:rq3-communication-expert-role}, Go Fan introduces the agent as a
Go module expert responsible for reviewing dependencies. This specifies the expertise
expected for the task, beyond identifying it as an AI agent.

\begin{rq3mdexample}{Role Assignment}{github/gh-aw}{go-fan.md}{https://github.com/github/gh-aw/blob/1170fc484978ff65b6e026755866aea0af490710/.github/workflows/go-fan.md\#L57-L57}{snip:rq3-communication-expert-role}
You are the \textbf{Go Fan} - an enthusiastic Go module expert who performs daily deep reviews of the Go dependencies used in this project. Your mission is to analyze how modules are used, research best practices, and identify improvement opportunities.\par
\end{rq3mdexample}

\noindent \rqsubcat{Automation Disclosure} (25.0\%) requires outputs to
identify automated authorship or disclose limitations, uncertainty, or missing
checks. For instance, in Snippet~\ref{snip:rq3-communication-automation-disclosure},
Handbookbot must identify itself to contributors as an automated assistant. The
disclosure belongs in the response, not just in the workflow's internal role
description.

\begin{rq3mdexample}{Automation Disclosure}{NikiforovAll/claude-code-rules}{handbookbot.md}{https://github.com/NikiforovAll/claude-code-rules/blob/79fbbd2187080066e54ac81cc22fb199dd81e1fc/.github/workflows/handbookbot.md\#L158-L158}{snip:rq3-communication-automation-disclosure}
\begin{itemize}[leftmargin=1.3em,itemsep=1pt,parsep=0pt,topsep=2pt]
\item \textbf{Transparent about your nature}: Always identify yourself as Handbookbot, an automated AI assistant.
\end{itemize}
\end{rq3mdexample}

\noindent \rqsubcat{Audience Voice} (20.1\%) specifies tone, audience, language,
or communication style, such as writing for maintainers, using polite language,
or adopting an encouraging tone. For example, one workflow instructs the agent to
\textit{``Be positive,
encouraging, and helpful''} and \textit{``Use emojis moderately for engagement''}.\rqexamplesource{https://github.com/github/gh-aw/blob/b5fdd698c629c91b387d059628db059bcda73980/.github/workflows/daily-team-status.md\#L45-L48}


\medskip
\noindent\faIcon{shield-alt}~\textbf{10 SAFETY (25.0\%).} Safety instructions
address untrusted input, restricted execution, sensitive information, and
security weaknesses. \rqsubcat{Execution Isolation} (13.9\%) specifies sandbox,
filesystem, network, permission, or read-only restrictions. For example, in
Snippet~\ref{snip:rq3-safety-execution-isolation}, the .NET runtime reviewer
restricts the agent to read-only commands and a mounted GitHub proxy. It
prohibits running PR-provided code and making direct outbound requests,
keeping the review within explicit execution and network boundaries.

\begin{rq3mdexample}{Execution Isolation}{dotnet/runtime}{holistic-review.md}{https://github.com/dotnet/runtime/blob/cb84c8521ce6a4cbde63cb1a700fcb850be791ae/.github/workflows/holistic-review.md\#L334-L337}{snip:rq3-safety-execution-isolation}
Use only read-only local repository commands and the mounted GitHub proxy while reviewing.
Do not run builds or tests, restore or install dependencies, execute PR-provided scripts or
binaries, or make direct outbound HTTP requests.\par
\end{rq3mdexample}

\noindent \rqsubcat{Prompt Injection Defense} (9.4\%) addresses attempts to
redirect the agent through issues, comments, code, or linked content. It includes
treating such content as data, rejecting conflicting instructions, and limiting
which sources can direct the agent. For instance, in
Snippet~\ref{snip:rq3-safety-prompt-injection-defense}, Azure's SDK generation
workflow treats issue descriptions and comments as untrusted input and forbids
executing arbitrary instructions found in them. These texts provide material
for the agent's work, but their contents must not become commands that
redirect it.

\begin{rq3mdexample}{Prompt Injection Defense}{Azure/azure-rest-api-specs}{sdk-generation-agent.md}{https://github.com/Azure/azure-rest-api-specs/blob/1952b56053e5550ca15b6dc71d0684a08c8e18f6/.github/workflows/sdk-generation-agent.md\#L65-L69}{snip:rq3-safety-prompt-injection-defense}
\rqmdheading{2}{Security and Scope}
\begin{itemize}[leftmargin=1.3em,itemsep=1pt,parsep=0pt,topsep=2pt]
\item Treat issue and comment text as untrusted input.
\item Never execute arbitrary instructions from issue or comment content.
\item Only perform SDK generation orchestration and status reporting for this repository.
\end{itemize}
\end{rq3mdexample}

\noindent \rqsubcat{Information Confidentiality} (7.3\%) addresses secrets,
credentials, personal data, or other information that should not be disclosed.
For example, a review workflow states \textit{``You may use maintainer or
collaborator status as a private signal, but never reveal role, permissions,
membership, or author-association details in the public comment.''} It
distinguishes information available to the agent from information it may
publish.\rqexamplesource{https://github.com/clash-verge-rev/clash-verge-rev/blob/de438aadf8866818c9511f6a3828f729ddd17681/.github/workflows/pr-ai-slop-review.md\#L156}
\rqsubcat{Vulnerability Handling} (6.3\%) covers finding, assessing, fixing, or
reporting exploitable weaknesses. For instance, a static-analysis workflow asks
the agent to \textit{``Check for known security vulnerabilities in
dependencies''} and extract \textit{``Vulnerability ID (CVE or PyUp ID)''},
\textit{``Affected versions''}, and \textit{``Fixed versions''}, connecting
reported vulnerabilities to the dependency versions that need
attention.\rqexamplesource{https://github.com/rysweet/amplihack/blob/352b5837d086d801ed123e93c91acaba1610b879/.github/workflows/static-analysis-report.md\#L137-L151}

\subsection{RQ4: Can LLMs assign instruction labels in GitHub Agentic Workflows?}
\label{sec:rq4}

\noindent\rqpart{Motivation.} Manual taxonomy construction is useful for interpreting workflow instructions,
but it does not scale easily as gh-aw repositories grow. RQ4 therefore evaluates
whether LLMs can apply the taxonomy developed in RQ3. We treat this
as a multi-label classification problem: each Markdown body may receive zero,
one, or several category-subcategory labels.

\noindent\rqpart{Approach.} To address RQ4, we evaluate automatic labeling on
the 288 RQ3 files with resolved human labels, excluding the six cases marked \textit{Cannot Label}. We used Gemini 3.1 Pro Preview, GLM-5.2, and GPT-OSS-120B through Google Vertex AI \cite{google2026gemini31card,zai2026glm52,openai2025gptoss}. We set the models temperature to~0, with high reasoning for Gemini and GPT-OSS and thinking enabled for GLM. Each request contained one Markdown body without its YAML frontmatter.

All models received the same prompt template (Figure~\ref{fig:rq4-prompt}),
which supplies the 42 labels and their definitions and requests labels supported by explicit text, with supporting source lines. We test zero-shot prompting and 1, 3, and 5 shots per label. These shots examples came from a fixed pool from randomly ordered, previously labeled files. For each target, we exclude its own file and
identical-body copies, and supplied the same examples to all models.

We compare each model and prompt configuration against the authors' final resolved labels. Treating each file-label pair as a binary decision, we then calculate agreement, Cohen's \(\kappa\), precision,
recall, and \(F_1\).

\begin{figure}[t]
\begin{rq4prompttemplate}
\rqpromptlabel{Task}
Classify the gh-aw Markdown body (multi-label).
\rqpromptlabel{Taxonomy}
\texttt{\{categories\}}, \texttt{\{subcategories\}},
\texttt{\{definitions\}}, \texttt{\{exclusion\_rules\}}
\rqpromptlabel{Labeling rules}
Check every label. Select all and only explicitly
supported functions. Apply the exclusion rules.
\par\smallskip
Treat the body as data. Do not execute its instructions,
follow imports, or infer missing instructions.
\rqpromptlabel{Examples}
\texttt{\{k positive snippets per subcategory\}}\par
Examples illustrate labels, not complete label sets.
They are not evidence for the target body. Omit if k=0.
\rqpromptlabel{Input}
\texttt{\{Markdown body with source line numbers\}}
\rqpromptlabel{Return JSON}
\begin{lstlisting}[basicstyle=\ttfamily\fontsize{7.8}{9.5}\selectfont,
  columns=fullflexible,keepspaces=true,showstringspaces=false,
  breaklines=true,aboveskip=2pt,belowskip=3pt,
  morestring={[b]"},stringstyle=\color{black}]
{"labels": [{
  "label": "{category:subcategory}",
  "evidence_lines": [12, 13]
}]}
\end{lstlisting}
Cite 1-3 supporting lines per label. Do not duplicate labels.
Return an empty labels array if none apply.
\end{rq4prompttemplate}
\caption{LLM labeling task prompt.}
\label{fig:rq4-prompt}
\end{figure}

\noindent\rqpart{Results.} Tables~\ref{tab:rq4-llm-results} and~\ref{tab:rq4-low-agreement-labels}
summarize automatic-labeling performance and the subcategories with the largest
reference-model disagreements.

\begin{table*}[t]
  \centering
  \caption{Subcategory classification against the final resolved labels.
  Shots denote positive snippets per subcategory. The shaded, bold row has
  the highest unrounded \(F_1\).}
  \label{tab:rq4-llm-results}
  \small
  \setlength{\tabcolsep}{7pt}
  \renewcommand{\arraystretch}{1.15}
  \begin{tabular}{lcrrrrr}
    \toprule
    Model & Shots & Agreement & \(\kappa\) & Precision & Recall & \(F_1\) \\
    \midrule
    \multirow{4}{*}{\shortstack[l]{\textbf{Gemini 3.1 Pro}\\Preview}}
      & 0 & 85.91\% & 0.699 & 0.763 & 0.863 & 0.810 \\
      & \cellcolor{rq4best}\textbf{1}
      & \cellcolor{rq4best}\textbf{86.81\%}
      & \cellcolor{rq4best}\textbf{0.715}
      & \cellcolor{rq4best}\textbf{0.786}
      & \cellcolor{rq4best}\textbf{0.853}
      & \cellcolor{rq4best}\textbf{0.818} \\
      & 3 & 86.81\% & 0.715 & 0.787 & 0.852 & 0.818 \\
      & 5 & 86.67\% & 0.711 & 0.788 & 0.844 & 0.815 \\
    \midrule
    \multirow{4}{*}{\textbf{GLM-5.2}}
      & 0 & 84.90\% & 0.680 & 0.741 & 0.869 & 0.800 \\
      & 1 & 86.00\% & 0.697 & 0.778 & 0.835 & 0.806 \\
      & 3 & 86.42\% & 0.704 & 0.790 & 0.830 & 0.810 \\
      & 5 & 86.30\% & 0.700 & 0.794 & 0.820 & 0.806 \\
    \midrule
    \multirow{4}{*}{\textbf{GPT-OSS-120B}}
      & 0 & 80.73\% & 0.575 & 0.724 & 0.720 & 0.722 \\
      & 1 & 81.56\% & 0.585 & 0.758 & 0.690 & 0.722 \\
      & 3 & 82.28\% & 0.602 & 0.766 & 0.706 & 0.735 \\
      & 5 & 82.41\% & 0.606 & 0.767 & 0.710 & 0.738 \\
    \bottomrule
  \end{tabular}
\end{table*}

\rqfinding{6}{One-shot Gemini reaches the highest observed \(F_1\)
(0.818), with 86.81\% subcategory agreement and Cohen's \(\kappa=0.715\).}
\nopagebreak[4]

Gemini has the highest \(F_1\) at every shot setting, followed by GLM and
GPT-OSS (Table~\ref{tab:rq4-llm-results}). One-shot and three-shot Gemini both
round to 0.818, with a slight advantage for one-shot before rounding.
GLM peaks at three shots (0.810), whereas GPT-OSS improves from 0.722
at zero shots to 0.738 at five shots. More examples therefore do not benefit
every model equally. These differences are descriptive, not evidence of
statistical superiority across repeated runs.

We inspect one-shot Gemini further because it has the highest observed
\(F_1\). Its 86.81\% agreement includes jointly absent labels; it assigns
15.86 subcategories per file, compared with 14.62 in the resolved reference.

\begin{table*}[t]
  \centering
  \caption{Lowest subcategory agreement for one-shot Gemini against the
  final resolved labels (288 files). Reference and Gemini count positive
  assignments; FP and FN denote false positives and false negatives.}
  \label{tab:rq4-low-agreement-labels}
  \small
  \setlength{\tabcolsep}{3.5pt}
  \begin{tabular}{lrrrrrr}
    \toprule
    Subcategory & Reference & Gemini & FP & FN & Agreement & \(\kappa\) \\
    \midrule
    Domain Context & 98 & 174 & 83 & 7 & 68.75\% & 0.414 \\
    Rule Conformance & 136 & 123 & 33 & 46 & 72.57\% & 0.447 \\
    Control Signal & 150 & 170 & 45 & 25 & 75.69\% & 0.510 \\
    \bottomrule
  \end{tabular}
\end{table*}

The largest disagreement concerns \textit{Domain Context}: Gemini assigns it
to 174 files, versus 98 in the resolved reference, with 83 false positives and
seven false negatives (Table~\ref{tab:rq4-low-agreement-labels}).
\textit{Rule Conformance} differs on 79 files, and \textit{Control Signal}
on 70. Domain Context's errors are predominantly additional assignments,
suggesting that Gemini sometimes reads background knowledge into an instruction.

For example, \texttt{check-emulator-updates.md} in
\texttt{drehelis/gcp-emulator-ui} asks the agent to
\textit{``Check the loaded text specifically for any mentions of''}
\texttt{emulator}, then lists services such as Pub/Sub, Datastore, and Spanner. Gemini cites
this line for \textit{Domain Context}, which the resolved labels omit.\footnote{\url{https://github.com/drehelis/gcp-emulator-ui/blob/744f45c95a03fdfcec7f0fabb3f547a3e0975b1e/.github/workflows/check-emulator-updates.md\#L51}}
The line specifies what to search for; it does not explain emulator behavior
or supply a domain rule. Gemini's assignment is consistent with treating
the technical concepts implied by the task as background knowledge.
This illustrates how an LLM may label an instruction's implied context as
well as its explicit function. It does not establish the model's internal
reasoning or show that every additional label is unsupported. The practical
boundary is between knowledge an instruction supplies and knowledge its
execution may require.

\section{Discussion and Implications}

In this section, we discuss implications for developers and researchers.

\subsection{For Developers}

\textbf{(1) Developers should monitor cross-project copied or referenced gh-aw
Markdown workflows as evolving dependencies, not as static templates.}
In our dataset, 108 of 365 projects (29.6\%) reference Markdown sources from
another repository for at least one generated workflow. Developers can
regenerate the YAML from these sources using \texttt{gh aw compile}.
RQ1 shows that these
Markdown sources are already substantial artifacts: the median body contains 104
lines and 556.5 words, and the median YAML frontmatter contains 58 lines. Among
RQ2's files with at least 120 days of activity, 78.2\% still receive updates
in month 4, despite a drop in size-normalized churn after the first month.
Moreover, one-third of content-changing file events in their repositories
leave the line count unchanged, so size alone cannot reveal changes to instructions.
Therefore, when a project copies or references a Markdown
workflow from another repository, developers should treat that source file like
an external workflow dependency. They should track upstream changes, pin or review the
commit used to generate their YAML, and recompile only after checking whether the
new instructions change permissions, tools, runtime cost, output behavior, or
agent authority.

\textbf{(2) Developers should consider how an agent carries out recurring work,
not only which tasks it executes.}
RQ3 identifies instruction choices beyond task descriptions that developers
can consider for their workflows. For example, resource budgets can set limits
on searches, retries, or token use to support cost control. Moreover, workflows
that read issues, comments, or online documents can include prompt-injection
defenses that distinguish external content from instructions to follow, as
illustrated in Snippet~\ref{snip:rq3-safety-prompt-injection-defense}.
For recurring work, cross-run memory can retain earlier findings or completed
actions, allowing the agent to build on previous runs rather than start from
scratch. These options should be selected for the workflow's needs. Our
taxonomy documents their use, but does not establish their effects on cost,
security, or repeated work.

\textbf{(3) Developers should distinguish rules enforced through frontmatter
from instructions the agent is asked to follow in the Markdown body.}
RQ2 shows that developers maintain both frontmatter and Markdown bodies, while RQ3 shows that body instructions specify how agents should work, not only their tasks. Some of these rules can also be enforced through frontmatter. For example, in \texttt{dotnet/aspnetcore}'s \texttt{triage-comment reviewer.md} \cite{aspnetcore2026triagecomment},
the body instructs the agent to ``post \textbf{at most one} comment'',
while the corresponding frontmatter settings limit the safe-output handler
to one comment:
\begin{lstlisting}[basicstyle=\ttfamily\small,columns=fullflexible,
  keepspaces=true,showstringspaces=false,frame=none,
  aboveskip=4pt,belowskip=4pt]
safe-outputs:
  add-comment:
    max: 1
\end{lstlisting}
Developers should use supported configuration controls for firm limits,
while using body instructions to explain task-specific conditions and
expectations. When a rule appears in both regions, they should keep the
two consistent as the workflow evolves.

\subsection{For Researchers}

\textbf{(1) Researchers should evaluate whether pictographic symbols make
agent-facing Markdown more efficient or more error-prone than plain text.}
RQ1 shows that 47.4\% of files contain at least one pictographic symbol, and the
largest case contains 171 symbols. That workflow uses check-mark and cross
symbols in its test-result table
\cite{github-gh-aw-mcpg-smoke-proxy-github-script}. Prior work
has studied prompt evolution, agent manifests, skills, and repository context
files \cite{tafreshipour2025prompting,chatlatanagulchai2025manifests,hong2026anatomy,gloaguen2026agentsmd},
but we did not find work that isolates whether symbol-heavy Markdown improves
agent task completion, reduces tokens, or introduces interpretation errors.
Future experiments should compare icon-based and text-based variants of the same
workflow under controlled task, model, and tokenizer settings.

\textbf{(2) Researchers should measure which instruction patterns actually help
agents, instead of treating all Markdown guidance as equally useful context.}
RQ3 shows substantial variation in how workflows guide agents: 89.9\% provide
resources, 68.1\% provide examples, 63.9\% include communication guidance, and
19.4\% include a worked case. These differences create a natural research
agenda for controlled ablations: remove or add resources, worked examples,
communication roles, output templates, and budget constraints, then measure task
completion, correctness, tool calls, runtime, and token usage. This is especially
important because recent work on repository-level context files shows that
additional agent-facing Markdown can increase exploration and inference cost by
more than 20\%, while producing mixed effects on task success
\cite{gloaguen2026agentsmd}.

\section{Threats to Validity}

\threatlabel{Internal Validity} Instruction types can be interpreted differently
by coders. To reduce this risk, two authors calibrated their coding on 30 files,
used shared definitions, assessed agreement before resolution, and reconciled
disagreements against the Markdown bodies. For automatic labeling, examples from
the target could reveal its assignments. We excluded the target and identical-body
copies from its examples and kept the prompt and examples consistent across models.

\threatlabel{Construct Validity} FRE is designed for prose, whereas our full-body
scores include code, Markdown syntax, and template expressions. We therefore
report these as descriptive formula outputs, not validated measures of human
or agent comprehension, and document the input text and preprocessing.
For instruction classification, mentioning a technical concept does not necessarily
supply domain context, as illustrated by the RQ4 disagreement example. We required
explicit textual support for labels rather than inferring instructions from
terminology or frontmatter. Similarly, unchanged file size does not imply unchanged
instructions, as the RQ2 command-replacement example shows. We compared consecutive
snapshots to measure additions, deletions, and affected regions rather than relying
on net size changes. Events whose regions could not be reliably compared were
left unclassified.

\threatlabel{Conclusion Validity} Repository activity declined during our
observation window, so fewer Markdown commits alone would not establish a
workflow-specific slowdown. We compared maintenance commits with overall
repository commits in the same months. Moreover, we used non-cumulative activity
and size-normalized churn to distinguish monthly editing from accumulated changes
and file growth. Statistical tests paired projects rather than treating their
files as independent, with Holm correction for multiple comparisons.
For automatic labeling, agreement includes labels absent from both predictions
and human annotations. We therefore also reported Cohen's \(\kappa\), precision,
recall, and \(F_1\), and inspected the largest label disagreements. We describe
model rankings as observed results, not statistical superiority across repeated
runs. GPT-OSS recovery requests changed generation settings, so the comparison
concerns the tested configurations rather than an identical reasoning budget.

\threatlabel{External Validity} GitHub's search limit could restrict corpus
coverage. We split searches into size buckets and retrieved all pages from buckets
below that limit, then randomly sampled collected Markdown files for RQ3.
These steps reduce missing search results and bias from choosing particular files,
but do not establish generalizability beyond our selection criteria.
In particular, RQ2 uses the 43 projects in our dataset containing files whose
observed changes span at least 120 days. This requirement supports comparison
over a common period but selects already-maintained files. Our findings therefore
describe their maintenance patterns, not how often all gh-aw workflows remain
static after creation.

\section{Related Work}
\label{sec:related-work}

\subsection{GitHub Actions Workflows}

Existing research on GitHub Actions examines how projects adopt, configure, and maintain
repository automation. For instance, Decan et al.\ \cite{decan2022actions} study workflow
usage across 68,000 repositories and show that reuse is concentrated in a
small set of actions. With regard to workflow maintenance, Valenzuela-Toledo et al.\
\cite{valenzuela2024hidden} identify bug fixing and CI/CD improvements as major
drivers of workflow changes. More recently, Rostami Mazrae et al.\
\cite{rostami2026githubactions} find that workflow edits are usually small and
mainly concern task specification and configuration. Together, these studies
show that YAML workflows require continued maintenance even though they
automate repetitive work. Moreover, Studies on GitHub Actions investigated AI-based automation. For example, Sun et al.\
\cite{sun2026review} examine 16 AI code-review actions and found that concise
comments containing code snippets are more likely to lead to changes.
Focusing on CI reliability, Chouchen et al.\ \cite{chouchen2026reliability}
analyze 11,771 agent- and human-authored pull requests to examine CI failures,
repair responsibility, and time to fix. They find that agent-authored fixes
are faster, but human developers remain involved in resolving failures. Furthermore, Ghaleb \cite{ghaleb2026configurations} examines agents edits
to CI/CD configurations, finding that these edits are uncommon and mainly
target GitHub Actions. Ghaleb and Rathnayake \cite{ghaleb2025writeci}
evaluate LLMs generating GitHub Actions YAML from natural-language descriptions,
identifying missing steps and misinterpreted requirements. Extending this
line of work, Ghaleb's Doc2CI study \cite{ghaleb2026doc2ci} shows that generated
configurations can parse as YAML yet violate a CI service's schema, and
evaluates schema-guided repair.

These studies examine AI-generated configurations, workflow maintenance, and
agents' contributions to CI. However, using an agent to create or edit a YAML
pipeline differs from defining a workflow whose recurring task is carried out
by an agent. We study the Markdown sources that specify such tasks in gh-aw,
examining their structure, maintenance, and instruction content. In particular,
we distinguish changes to execution configuration from changes to the
instructions that guide the agent.

\subsection{Agent Instructions and Context Files}

In today's agentic AI era, natural-language instructions such as skills.md are themselves maintained as software artifacts.
Tafreshipour et al.\ \cite{tafreshipour2025prompting} study prompt evolution
in software repositories, finding that additions and modifications commonly
accompany feature development and are often poorly documented. For
agent-facing Markdown, Chatlatanagulchai et al.\
\cite{chatlatanagulchai2025manifests} identify build commands, implementation
details, and architecture as common contents of \texttt{CLAUDE.md} files.
Their broader study of agent context files finds frequent, small additions
and an emphasis on functional guidance over security and performance
requirements \cite{chatlatanagulchai2026agentreadmes}. Similarly,
Mohsenimofidi et al.\ \cite{mohsenimofidi2026context} report varied structures
and ways of expressing instructions in \texttt{AGENTS.md} files.

Beyond general project context, Hong et al.\ \cite{hong2026anatomy} derive a
taxonomy of \texttt{SKILL.md} contents and identify violations of skill-authoring
recommendations. Thus, existing research already covers procedural knowledge,
not only descriptive documentation. Studies also test whether instructions
benefit agents. Lulla et al.\ \cite{lulla2026efficiency} associate
\texttt{AGENTS.md} with lower runtime and output-token consumption in their
experiment. In a different evaluation, Gloaguen et al.\
\cite{gloaguen2026agentsmd} find that context files do not generally improve
task success and increase inference costs. These results caution against
treating the presence of instructions as evidence of their effectiveness.

Valenzuela-Toledo et al.\ \cite{valenzuela2026ghawhistories} recently released
GHAW-H, a dataset linking gh-aw Markdown histories to compiled YAML snapshots. Our study
provides complementary empirical evidence through a separately collected
dataset, a maintenance analysis, a manually developed instruction taxonomy,
and an evaluation of automatic labeling.

The distinction from general context and skill files is therefore not the
use of Markdown or the presence of procedures. In gh-aw, instructions and
execution configuration together define a workflow invoked by repository
events or schedules \cite{github-gh-aw-overview,github-gh-aw-compilation}.
We examine how developers maintain these combined specifications and which
instruction types they use for recurring work. Our taxonomy describes those
choices, while our classification experiment evaluates whether LLMs can
identify them, not whether the instructions improve execution outcomes.

\section{Conclusion}

This study examined gh-aw Markdown files as software artifacts that
combine executable workflow configuration with natural-language
instructions for agents. Our analysis of 1,248 files shows that these
artifacts contain substantial, structured specifications: their
instruction bodies combine prose with lists, code blocks, templates,
and headings, while their frontmatter defines execution settings
such as triggers, tools, and permissions. They also require continued
maintenance. Among the 348 files with at least 120 days of observed
activity, 78.2\% received updates in the fourth month. Although
size-normalized churn declined after the first month, the share
of repository commits devoted to workflow maintenance did not
change significantly between adjacent months. Changes affected
both configuration and instructions, with instruction bodies
accounting for most changed lines in the analyzed workflow files.

Our taxonomy shows that developers express a broad range of
operational requirements in these instructions, including tasks,
expected outputs, constraints, procedures, context, and resources.
Safety instructions and resource budgets are less frequently
explicit, identifying areas for closer review without establishing
that corresponding runtime protections are absent.
Together, these findings show that adopting natural language for
workflow specification introduces an additional body of maintained
instructions whose content deserves explicit attention during
workflow review.

Future work should investigate how instruction patterns and
revisions affect execution outcomes, including task success,
reliability, cost, and safety. These evaluations can inform
validation and regression-testing techniques for agent instructions,
as well as guidance on when to modularize complex workflows.
Our dataset and taxonomy provide a foundation for this work by
characterizing the specifications developers currently write
and maintain.

\section*{Acknowledgment}
This work was supported by the Fonds de recherche du Québec – Nature et technologies (FRQNT) and the Natural Sciences and Engineering Research Council of Canada (NSERC). Additionally, we thank the Google Research Credits program for providing credits to run the RQ4 experiments on Vertex AI.

We declare that AI solutions (i.e., Grammarly and OpenAI's Codex) were used for grammar checking and rephrasing. We declare
that the content of this article and the whole research procedure were developed and fully conducted by the authors, taking full responsibility for the publication’s content.


\balance
\bibliographystyle{IEEEtran}
\begingroup
\setlength{\emergencystretch}{1em}
\bibliography{references}

@inproceedings{khelifi2025ghaminer,
  title={GHAminer: An open source tool to extract GitHub actions build metrics},
  author={Khelifi, Jasem and Benzina, Yacine and Chouchen, Moataz and Ouni, Ali and Sayagh, Mohammed and Bouktif, Salah},
  booktitle={IEEE International Conference on Software Analysis, Evolution and Reengineering (SANER)},
  pages={834--838},
  year={2025}
}

@article{khelifi2026continuous,
  title={Continuous Integration Workflows in Microservice-based Systems: Performance, Failures, and Recovery},
  author={Khelifi, Jasem and Ouni, Ali and Sayagh, Mohammed and Saied, Mohamed Aymen and Khelifi, Syrine and Bouktif, Salah},
  journal={ACM Transactions on Software Engineering and Methodology},
  year={2026}
}

@misc{runtime2026scanner,
  author = {{.NET Contributors}},
  title = {{CI Outer-Loop Failure Scanner}},
  year = {2026},
  howpublished = {\url{https://github.com/dotnet/runtime/blob/8f769f811a4b5b69328c175c9e2ae5d09cdc7c23/.github/workflows/ci-failure-scan.md}},
  note = {Commit-pinned workflow snapshot. Accessed 2026-09-15}
}

@misc{runtime2026fixer,
  author = {{.NET Contributors}},
  title = {{CI Outer-Loop Failure Fixer}},
  year = {2026},
  howpublished = {\url{https://github.com/dotnet/runtime/blob/8f769f811a4b5b69328c175c9e2ae5d09cdc7c23/.github/workflows/ci-failure-fix.md}},
  note = {Commit-pinned workflow snapshot. Accessed 2026-09-15}
}

@misc{runtime2026feedback,
  author = {{.NET Contributors}},
  title = {{CI Outer-Loop Failure Scanner: Feedback}},
  year = {2026},
  howpublished = {\url{https://github.com/dotnet/runtime/blob/8f769f811a4b5b69328c175c9e2ae5d09cdc7c23/.github/workflows/ci-failure-scan-feedback.md}},
  note = {Commit-pinned workflow snapshot. Accessed 2026-09-15}
}

@misc{runtime2026feedbackissue,
  author = {{.NET Contributors}},
  title = {{Route platform-specific crypto-config failures to loop-in}},
  year = {2026},
  howpublished = {\url{https://github.com/dotnet/runtime/issues/131037}},
  note = {Feedback workflow output, issue 131037. Accessed 2026-09-15}
}

@misc{runtime2026feedbackfix,
  author = {{.NET Contributors}},
  title = {{Route platform-specific crypto-config failures to loop-in}},
  year = {2026},
  howpublished = {\url{https://github.com/dotnet/runtime/pull/131469}},
  note = {Pull request 131469, merged July 29, 2026. Accessed 2026-09-15}
}

@misc{pichai2026cloudnext,
  author = {Pichai, Sundar},
  title = {{Cloud Next '26: Momentum and innovation at Google scale}},
  year = {2026},
  howpublished = {\url{https://blog.google/innovation-and-ai/infrastructure-and-cloud/google-cloud/cloud-next-2026-sundar-pichai/}},
  note = {Google Blog, April 22, 2026. Accessed 2026-09-15}
}

@misc{carlini2026compiler,
  author = {Carlini, Nicholas},
  title = {{Building a C compiler with a team of parallel Claudes}},
  year = {2026},
  howpublished = {\url{https://www.anthropic.com/engineering/building-c-compiler}},
  note = {Anthropic Engineering, February 5, 2026. Accessed 2026-09-15}
}

@misc{github-gh-aw-overview,
  author = {{GitHub}},
  title = {{About GitHub Agentic Workflows}},
  year = {2026},
  howpublished = {\url{https://docs.github.com/en/copilot/concepts/agents/about-github-agentic-workflows}},
  note = {GitHub Docs. Accessed 2026-09-15}
}

@inproceedings{yang2024sweagent,
  author = {Yang, John and Jimenez, Carlos E. and Wettig, Alexander and Lieret, Kilian and Yao, Shunyu and Narasimhan, Karthik and Press, Ofir},
  title = {{SWE-agent: Agent-Computer Interfaces Enable Automated Software Engineering}},
  booktitle = {Advances in Neural Information Processing Systems},
  editor = {Globerson, A. and Mackey, L. and Belgrave, D. and Fan, A. and Paquet, U. and Tomczak, J. and Zhang, C.},
  volume = {37},
  pages = {50528--50652},
  publisher = {Curran Associates, Inc.},
  year = {2024},
  doi = {10.52202/079017-1601},
  url = {https://proceedings.neurips.cc/paper_files/paper/2024/hash/5a7c947568c1b1328ccc5230172e1e7c-Abstract-Conference.html}
}

@inproceedings{valenzuela2024hidden,
  author = {Pablo Valenzuela-Toledo and Alexandre Bergel and Timo Kehrer and Oscar Nierstrasz},
  title = {The Hidden Costs of Automation: An Empirical Study on {GitHub Actions} Workflow Maintenance},
  booktitle = {International Conference on Source Code Analysis and Manipulation},
  year = {2024},
  pages = {213--223},
  doi = {10.1109/SCAM63643.2024.00029}
}

@misc{github-rest-search,
  author = {{GitHub}},
  title = {{REST API endpoints for search}},
  year = {2026},
  howpublished = {\url{https://docs.github.com/en/rest/search/search\#search-code}},
  note = {Search code. Accessed 2026-09-11}
}

@misc{github-rest-commits,
  author = {{GitHub}},
  title = {{REST API endpoints for commits}},
  year = {2026},
  howpublished = {\url{https://docs.github.com/en/rest/commits/commits}},
  note = {Accessed 2026-08-09}
}

@misc{github-gh-aw-mcpg-smoke-proxy-github-script,
  author = {{GitHub}},
  title = {{gh-aw-mcpg smoke-proxy-github-script workflow Markdown snapshot}},
  year = {2026},
  howpublished = {\url{https://github.com/github/gh-aw-mcpg/blob/45a6fb81de50cc5e4c7dd0232522492a74ca8160/.github/workflows/smoke-proxy-github-script.md}},
  note = {Accessed 2026-08-13}
}

@article{holm1979simple,
  author = {Holm, Sture},
  title = {{A Simple Sequentially Rejective Multiple Test Procedure}},
  journal = {Scandinavian Journal of Statistics},
  volume = {6},
  number = {2},
  pages = {65--70},
  year = {1979},
  doi = {10.2307/4615733}
}

@article{landis1977measurement,
  author = {Landis, J. Richard and Koch, Gary G.},
  title = {{The Measurement of Observer Agreement for Categorical Data}},
  journal = {Biometrics},
  volume = {33},
  number = {1},
  pages = {159--174},
  year = {1977},
  doi = {10.2307/2529310}
}

@misc{robbes2026agentic,
  author = {Robbes, Romain and Matricon, Th\'{e}o and Degueule, Thomas and Hora, Andre and Zacchiroli, Stefano},
  title = {{Agentic Much? Adoption of Coding Agents on GitHub}},
  year = {2026},
  howpublished = {arXiv:2601.18341},
  doi = {10.48550/arXiv.2601.18341}
}

@misc{github2026agentic-preview,
  author = {Syme, Don and de Halleux, Peli},
  title = {{Automate repository tasks with GitHub Agentic Workflows}},
  year = {2026},
  howpublished = {\url{https://github.blog/ai-and-ml/automate-repository-tasks-with-github-agentic-workflows/}},
  note = {GitHub Blog, February 13, 2026. Accessed 2026-09-10}
}

@misc{aspnetcore2026triagecomment,
  author = {{.NET Contributors}},
  title = {{triage-comment-reviewer.md: Triage Comment Reviewer for dotnet/aspnetcore}},
  year = {2026},
  howpublished = {\url{https://github.com/dotnet/aspnetcore/blob/5a09c126653d5bd295262f2ca96c9388b854cb24/.github/workflows/triage-comment-reviewer.md}},
  note = {Collected commit-pinned workflow snapshot}
}

@misc{github-gh-aw-compilation,
  author = {{GitHub}},
  title = {{Compilation Process}},
  year = {2026},
  howpublished = {\url{https://github.github.com/gh-aw/reference/compilation-process/}},
  note = {GitHub Agentic Workflows documentation. Accessed 2026-09-10}
}

@misc{github-gh-aw-repository,
  author = {{GitHub}},
  title = {{GitHub Agentic Workflows (gh-aw) repository}},
  year = {2026},
  howpublished = {\url{https://github.com/github/gh-aw}},
  note = {Accessed 2026-09-08}
}

@misc{activiti-supply-chain-lock,
  author = {{Activiti contributors}},
  title = {{supply-chain-review.lock.yml: Generated gh-aw workflow}},
  year = {2026},
  howpublished = {\url{https://github.com/Activiti/Activiti/blob/11e0c28347ca68e37848ae50710829e94488be7c/.github/workflows/supply-chain-review.lock.yml}},
  note = {Pinned snapshot. Accessed 2026-09-13}
}

@misc{awesome-copilot-duplicate-lock,
  author = {{GitHub awesome-copilot contributors}},
  title = {{duplicate-resource-detector.lock.yml: Generated gh-aw workflow}},
  year = {2026},
  howpublished = {\url{https://github.com/github/awesome-copilot/blob/4b6430ee3ed78c4029bbb7360ba117b5ce53a78d/.github/workflows/duplicate-resource-detector.lock.yml}},
  note = {Pinned snapshot. Accessed 2026-09-13}
}

@misc{pydantic-attention-triage-lock,
  author = {{Pydantic contributors}},
  title = {{pydantic-ai-attention-triage.lock.yml: Generated gh-aw workflow}},
  year = {2026},
  howpublished = {\url{https://github.com/pydantic/pydantic-ai/blob/a1defdf9e078bbb6cce979a98c931b342f7e05d8/.github/workflows/pydantic-ai-attention-triage.lock.yml}},
  note = {Pinned snapshot. Accessed 2026-09-13}
}

@misc{markdown-workflows-replication,
  author = {Khelifi, Jasem and Oukhay, Issam and Ouni, Ali and Sayagh Mohammed and Saied, Mohamed Aymen},
  title = {{ghaw replication package}},
  year = {2026},
  howpublished = {\url{https://github.com/stilab-ets/ghaw}},
  note = {Accessed 2026-08-13}
}

@incollection{braun2012thematic,
  author = {Braun, Virginia and Clarke, Victoria},
  title = {Thematic Analysis},
  booktitle = {APA Handbook of Research Methods in Psychology, Vol. 2: Research Designs: Quantitative, Qualitative, Neuropsychological, and Biological},
  editor = {Cooper, Harris and Camic, Paul M. and Long, Debra L. and Panter, A. T. and Rindskopf, David and Sher, Kenneth J.},
  publisher = {American Psychological Association},
  year = {2012},
  pages = {57--71},
  doi = {10.1037/13620-004}
}

@inproceedings{chouchen2024chatgpt,
  author = {Chouchen, Moataz and Bessghaier, Narjes and Begoug, Mahi and Ouni, Ali and AlOmar, Eman Abdullah and Mkaouer, Mohamed Wiem},
  title = {How Do Software Developers Use {ChatGPT}? An Exploratory Study on {GitHub} Pull Requests},
  booktitle = {International Conference on Mining Software Repositories},
  year = {2024},
  pages = {212--216},
  publisher = {ACM},
  doi = {10.1145/3643991.3645084}
}

@misc{hong2026anatomy,
  author = {Hong, David Boram and Imani, Aaron and Ahmed, Iftekhar},
  title = {{From Anatomy to Smells: An Empirical Study of SKILL.md in Agent Skills}},
  year = {2026},
  howpublished = {arXiv:2607.01456},
  doi = {10.48550/arXiv.2607.01456}
}

@inproceedings{tafreshipour2025prompting,
  author = {Tafreshipour, Mahan and Imani, Aaron and Huang, Eric and Almeida, Eduardo and Zimmermann, Thomas and Ahmed, Iftekhar},
  title = {{Prompting in the Wild: An Empirical Study of Prompt Evolution in Software Repositories}},
  booktitle = {Proceedings of the 22nd IEEE/ACM International Conference on Mining Software Repositories},
  year = {2025},
  doi = {10.1109/MSR66628.2025.00106}
}

@inproceedings{chatlatanagulchai2025manifests,
  author = {Chatlatanagulchai, Worawalan and Thonglek, Kundjanasith and Reid, Brittany and Kashiwa, Yutaro and Leelaprute, Pattara and Rungsawang, Arnon and Manaskasemsak, Bundit and Iida, Hajimu},
  title = {{On the Use of Agentic Coding Manifests: An Empirical Study of Claude Code}},
  booktitle = {International Conference on Product-Focused Software Process Improvement},
  year = {2025},
  doi = {10.1007/978-3-032-12089-2_40}
}

@misc{chatlatanagulchai2026agentreadmes,
  author = {Chatlatanagulchai, Worawalan and Li, Hao and Kashiwa, Yutaro and Reid, Brittany and Thonglek, Kundjanasith and Leelaprute, Pattara and Rungsawang, Arnon and Manaskasemsak, Bundit and Adams, Bram and Hassan, Ahmed E. and Iida, Hajimu},
  title = {{Agent READMEs: An Empirical Study of Context Files for Agentic Coding}},
  year = {2026},
  howpublished = {arXiv:2511.12884v2},
  doi = {10.48550/arXiv.2511.12884},
  url = {https://arxiv.org/abs/2511.12884v2},
  note = {Version 2, revised August 9, 2026}
}

@article{rostami2026githubactions,
  author = {Rostami Mazrae, Pooya and Decan, Alexandre and Mens, Tom and Wessel, Mairieli},
  title = {{An Empirical Study of the Evolution of GitHub Actions Workflows}},
  journal = {Journal of Systems and Software},
  volume = {236},
  pages = {112824},
  year = {2026},
  doi = {10.1016/j.jss.2026.112824}
}

@article{gloaguen2026agentsmd,
  author = {Gloaguen, Thibaud and Mundler, Niels and Muller, Mark and Raychev, Veselin and Vechev, Martin},
  title = {{Evaluating AGENTS.md: Are Repository-Level Context Files Helpful for Coding Agents?}},
  journal = {arXiv preprint arXiv:2602.11988},
  year = {2026}
}

@article{cohen1960kappa,
  author = {Cohen, Jacob},
  title = {A Coefficient of Agreement for Nominal Scales},
  journal = {Educational and Psychological Measurement},
  volume = {20},
  number = {1},
  pages = {37--46},
  year = {1960},
  doi = {10.1177/001316446002000104}
}

@article{eleyan2020enhancing,
  author = {Eleyan, Derar and Othman, Abed and Eleyan, Amna},
  title = {Enhancing Software Comments Readability Using Flesch Reading Ease Score},
  journal = {Information},
  volume = {11},
  number = {9},
  pages = {430},
  year = {2020},
  doi = {10.3390/info11090430}
}

@article{wouters2022crowd,
  author = {Wouters, Jelle and Menkveld, Abel and Brinkkemper, Sjaak and Dalpiaz, Fabiano},
  title = {Crowd-Based Requirements Elicitation via Pull Feedback: Method and Case Studies},
  journal = {Requirements Engineering},
  volume = {27},
  pages = {429--455},
  year = {2022},
  doi = {10.1007/s00766-022-00384-6}
}

@incollection{bala2024chatgpt,
  author = {Bala, S. and Sahling, K. and Haase, J. and Mendling, J.},
  title = {{ChatGPT} for Tailoring Software Documentation for Managers and Developers},
  booktitle = {Agile Processes in Software Engineering and Extreme Programming -- Workshops},
  series = {Lecture Notes in Business Information Processing},
  volume = {524},
  pages = {103--109},
  publisher = {Springer},
  year = {2024},
  doi = {10.1007/978-3-031-72781-8_11}
}

@article{prechelt2025abstracts,
  author = {Prechelt, Lutz and Montgomery, Lucy and Frattini, Julian and Zieris, Franz},
  title = {How (Not) To Write a Software Engineering Abstract},
  journal = {arXiv preprint arXiv:2506.21634},
  year = {2025}
}

@article{ehsani2026chatgpt,
  author = {Ehsani, R. and Pathak, S. and Parra, E. and Haiduc, S. and Chatterjee, P.},
  title = {What Characteristics Make {ChatGPT} Effective for Software Issue Resolution? An Empirical Study of Task, Project, and Conversational Signals in {GitHub} Issues},
  journal = {Empirical Software Engineering},
  volume = {31},
  number = {22},
  year = {2026},
  doi = {10.1007/s10664-025-10745-8}
}

@article{watanabe2026agenticprs,
  author = {Watanabe, Miku and Li, Hao and Kashiwa, Yutaro and Reid, Brittany and Iida, Hajimu and Hassan, Ahmed E.},
  title = {{On the Use of Agentic Coding: An Empirical Study of Pull Requests on GitHub}},
  journal = {ACM Transactions on Software Engineering and Methodology},
  year = {2026},
  doi = {10.1145/3798166},
  url = {https://arxiv.org/abs/2509.14745}
}

@inproceedings{he2026cursor,
  author = {He, Hao and Miller, Courtney and Agarwal, Shyam and K{\"a}stner, Christian and Vasilescu, Bogdan},
  title = {{Speed at the Cost of Quality: How Cursor AI Increases Short-Term Velocity and Long-Term Complexity in Open-Source Projects}},
  booktitle = {International Conference on Mining Software Repositories},
  pages = {181--193},
  year = {2026}
}

@article{salzano2026bridging,
  author = {Salzano, Francesco and Marchesi, Lodovica and Antenucci, Cosmo Kevin and Scalabrino, Simone and Tonelli, Roberto and Oliveto, Rocco and Pareschi, Remo},
  title = {Bridging the gap: a comparative study of academic and developer approaches to smart contract vulnerabilities},
  journal = {Empirical Software Engineering},
  volume = {31},
  number = {2},
  articleno = {37},
  year = {2026},
  doi = {10.1007/s10664-025-10780-5}
}

@misc{danilak2021langdetect,
  author = {Danilak, Michal Mimino},
  title = {{langdetect}: Language Detection Library for {Python}},
  year = {2021},
  howpublished = {\url{https://github.com/Mimino666/langdetect}},
  note = {Version 1.0.9. Accessed 2026-09-11}
}

@misc{google2026gemini31card,
  author = {{Google DeepMind}},
  title = {{Gemini 3.1 Pro}: Model Card},
  year = {2026},
  url = {https://deepmind.google/models/model-cards/gemini-3-1-pro/},
  note = {Accessed September 14, 2026}
}

@misc{zai2026glm52,
  author = {{Z.ai}},
  title = {{GLM-5.2}: Model Card},
  year = {2026},
  url = {https://huggingface.co/zai-org/GLM-5.2},
  note = {Accessed September 14, 2026}
}

@misc{openai2025gptoss,
  author = {{OpenAI}},
  title = {Introducing {gpt-oss}},
  year = {2025},
  url = {https://openai.com/index/introducing-gpt-oss/},
  note = {Accessed September 14, 2026}
}

@inproceedings{nagappan2005relative,
  author = {Nagappan, Nachiappan and Ball, Thomas},
  title = {Use of Relative Code Churn Measures to Predict System Defect Density},
  booktitle = {International Conference on Software Engineering},
  year = {2005},
  pages = {284--292},
  url = {https://www.microsoft.com/en-us/research/wp-content/uploads/2016/02/icse05churn.pdf}
}

@misc{nist2023signtest,
  author = {{National Institute of Standards and Technology}},
  title = {Sign Test},
  year = {2023},
  howpublished = {\url{https://www.itl.nist.gov/div898/software/dataplot/refman1/auxillar/signtest.htm}},
  note = {Dataplot reference manual. Accessed 2026-09-11}
}

@inproceedings{decan2022actions,
  author = {Decan, Alexandre and Mens, Tom and Rostami Mazrae, Pooya and Golzadeh, Mehdi},
  title = {On the Use of {GitHub Actions} in Software Development Repositories},
  booktitle = {International Conference on Software Maintenance and Evolution},
  year = {2022},
  pages = {235--245},
  doi = {10.1109/ICSME55016.2022.00029}
}

@article{sun2026review,
  author = {Sun, Kexin and Kuang, Hongyu and Baltes, Sebastian and Zhou, Xin and Zhang, He and Ma, Xiaoxing and Rong, Guoping and Shao, Dong and Treude, Christoph},
  title = {Does {AI} Code Review Lead to Code Changes? A Case Study of {GitHub Actions}},
  journal = {IEEE Transactions on Software Engineering},
  volume = {52},
  number = {7},
  pages = {2111--2126},
  year = {2026},
  doi = {10.1109/TSE.2026.3688237}
}

@inproceedings{chouchen2026reliability,
  author = {Chouchen, Moataz and Khelifi, Jasem and Begoug, Mahi and Ouni, Ali and Sayagh, Mohammed and Saied, Mohamed Aymen},
  title = {On the Reliability of Agentic {AI} in Continuous Integration Pipelines},
  booktitle = {Proceedings of the 23rd International Conference on Mining Software Repositories},
  publisher = {ACM},
  pages = {837--841},
  year = {2026},
  doi = {10.1145/3793302.3793585}
}

@inproceedings{ghaleb2026configurations,
  author = {Ghaleb, Taher A.},
  title = {When {AI} Agents Touch {CI/CD} Configurations: Frequency and Success},
  booktitle = {International Conference on Mining Software Repositories},
  publisher = {ACM},
  year = {2026},
  pages = {817--821},
  doi = {10.1145/3793302.3793581}
}

@inproceedings{ghaleb2025writeci,
  author = {Ghaleb, Taher A. and Rathnayake, Dulina},
  title = {Can {LLMs} Write {CI}? A Study on Automatic Generation of {GitHub Actions} Configurations},
  booktitle = {2025 IEEE International Conference on Software Maintenance and Evolution (ICSME)},
  publisher = {IEEE},
  year = {2025},
  pages = {767--772},
  doi = {10.1109/ICSME64153.2025.00077}
}

@misc{ghaleb2026doc2ci,
  author = {Ghaleb, Taher A.},
  title = {{Doc2CI}: A Multi-Service Study of {CI} Configuration Generation Using Large Language Models},
  year = {2026},
  howpublished = {arXiv:2608.01451},
  doi = {10.48550/arXiv.2608.01451}
}

@inproceedings{mohsenimofidi2026context,
  author = {Mohsenimofidi, Seyedmoein and Galster, Matthias and Treude, Christoph and Baltes, Sebastian},
  title = {Context Engineering for {AI} Agents in Open-Source Software},
  booktitle = {Proceedings of the 23rd International Conference on Mining Software Repositories},
  publisher = {ACM},
  year = {2026},
  pages = {194--198},
  doi = {10.1145/3793302.3793350}
}

@misc{lulla2026efficiency,
  author = {Lulla, Jai Lal and Mohsenimofidi, Seyedmoein and Galster, Matthias and Zhang, Jie M. and Baltes, Sebastian and Treude, Christoph},
  title = {On the Impact of {AGENTS.md} Files on the Efficiency of {AI} Coding Agents},
  year = {2026},
  howpublished = {arXiv:2601.20404}
}

@misc{valenzuela2026ghawhistories,
  author = {Valenzuela-Toledo, Pablo and Kehrer, Timo and Panichella, Sebastiano},
  title = {{GHAW-H}: A Dataset of {GitHub Agentic Workflow} Histories},
  year = {2026},
  howpublished = {Zenodo, version 0.1.2},
  doi = {10.5281/zenodo.22084012},
  url = {https://pavt.github.io/GHAW-H/assets/GHAW-H.pdf},
  note = {Dataset and companion report. Accessed September 18, 2026}
}

@article{johnson2025procedure,
  author = {Johnson, Karl and Morais, Caroline and Patelli, Edoardo},
  title = {Enhancing Procedure Quality: Advanced Language Tools for Identifying Ambiguity and High-Potential Violation Triggers},
  journal = {Reliability Engineering \& System Safety},
  year = {2025},
  volume = {264},
  pages = {111308},
  doi = {10.1016/j.ress.2025.111308}
}

@article{ali2022mapping,
  author = {Ali, Imran and Kannan, Devika},
  title = {Mapping Research on Healthcare Operations and Supply Chain Management: A Topic Modelling-Based Literature Review},
  journal = {Annals of Operations Research},
  year = {2022},
  volume = {315},
  number = {1},
  pages = {29--55},
  doi = {10.1007/s10479-022-04596-5}
}
\endgroup

\end{document}